\documentclass[12pt]{iopart}
\usepackage{xcolor, graphicx}
\usepackage[normalem]{ulem}
\usepackage{braket}
\usepackage[caption=false]{subfig}

\newcommand{\void}[1]{}
\newcommand{\be}{\begin{equation}}
\newcommand{\ee}{\end{equation}}

\newcommand{\non}{\nonumber}

\begin{document}
\title[Quench dynamics of interacting bosons]{Quench dynamics of interacting bosons: generalized coherent states versus multi-mode Glauber states}
\author{Yulong Qiao} 
\address{Institut f\"{u}r Theoretische Physik, Technische Universit\"{a}t Dresden, D-01062 Dresden, Germany}
\author{Frank Grossmann} 
\address{Institut f\"{u}r Theoretische Physik, Technische Universit\"{a}t Dresden, D-01062 Dresden, Germany}
\date{\today}

\begin{abstract}
Multi-mode Glauber coherent states (MMGS) as well as Bloch states with zero 
quasi-momentum, which are a special case of generalized coherent states (GCS), 
are frequently used to describe condensed phases of bosonic many-body systems. 
The difference of  two-point correlators of MMGS and GCS vanishes in the thermodynamic
limit. Using the established expansion of GCS in terms of MMGS, 
we derive a Fourier-type relation between the (auto-)correlation functions of the 
two different time-evolved states. This relation reveals that the (auto-)correlation and 
thus the dynamical free-energy density for the two cases are still different, even in the 
thermodynamic limit, due to the lack of the U(1) symmetry of the MMGS. Analytic 
results for the deep lattice model of interacting bosons for increasing filling 
factors show multiple sharp structures in the dynamical free energy-density of 
increasing complexity. These are explained using the evolution of Husimi 
functions in phase space. 
\end{abstract}

\maketitle

\section{Introduction}
The dynamics of many-body quantum systems after a quench, i.e., after a slow or sudden 
parameter change in the Hamiltonian is at the heart of a plethora of studies in condensed matter physics and quantum information \cite{Mitra18}. Also in cold atom physics, most
experiments are based on the propagation of an initial state after the trap parameters 
(and thus the Hamiltonian) are suddenly or adiabatically changed, see, e.g. \cite{GMHB02,PSSV11,Tetal12}.

In the seminal study of Greiner et al \cite{GMHB02}, the revival dynamics of 
a Bose-Einstein condensate in a deep optical lattice has been investigated experimentally
and described theoretically with a (multi-mode) Glauber coherent state
initial condition. It has been mentioned, however, that the MMCS is only
the thermodynamic limit of the true condensate ground state, described by the generalized coherent states (GCS) \cite{Pere71,Pere,ACGT72,ZFG90}, which is given by a 
product of identical single particle Bloch waves with zero quasi-momentum. Some static properties of GCS are similar to the MMCS', for example, the nonorthogonality and the impact of annihilation operators acting on them. More importantly, both states have off-diagonal long-range order, such that they are suitable to describe a Bose-Einstein condensate (BEC) \cite{Yang62,LSY07}. Nonetheless, there are distinctions between MMCS and GCS. A primary difference is that GCS maintains a definite number of particles, whereas MMCS exhibits fluctuations in particle number leading to U(1) symmetry breaking. These differences and similarities are reflected in their non-equilibrium features. The time-dependent dynamics of MMCS and GCS governed by two-body interactions has been compared in \cite{SDZ11}. They proved that when considering the thermodynamical limit, the MMCS is equivalent to the GCS in terms of the evolution of two-point correlation function. Although the discrepancy between MMCS and GCS is not manifested in the two-point correlation case, this work will highlight the dynamical difference stemming from U(1) symmetry by examining the evolution of the autocorrelation function. The dynamics of the autocorrelation function has drawn a lot of attention, because its periodic collapse and revival in some systems is the manifestation of quantum interference \cite{WaHe09}. More recently, the autocorrelation function is also used to study quantum dynamical phase transitions due to its analogy to the partition functions, and the transitions happens at instances of time, when the autocorrelation function is zero \cite{HPK13}.
\void{\textcolor{yellow}{The use of the "naive" MMCS initial state has again been studied in \cite{SDZ11}, where 
the necessity of finite  size (particle number) corrections has been pointed out. 
The initial states to be employed instead of the standard Glauber coherent states 
to theoretically account for finite particle numbers
were shown to be the generalized coherent states (GCS), pioneered by Perelomov 
\cite{Pere71,Pere} and  Gilmore and coworkers \cite{ACGT72,ZFG90}.}}

The use of GCS to treat the
quantum dynamics of finite bosonic lattice models of Bose-Hubbard type
has also been put forth more recently \cite{Wim21,pra21,front23}. In \cite{Wim21},
a mean-field variational approach, based on an ansatz for the solution of the time-dependent
Schr\"odinger equation in terms of a single GCS (termed atomic coherent state) was shown 
to correctly predict some finite size effects in the bosonic Josephson junction 
(double well Bose-Hubbard system), with unsatisfactory results for more subtle quantum  effects like  spontaneous symmetry breaking and macroscopic self trapping, however.
In \cite{front23}, a multi-configuration variational approach based on GCS  was taken 
to overcome  those deficiencies.  In contrast to the well-established multi-configuration 
time-dependent Hartree method  (MCTDH) \cite{BJWM00}, the basis functions used in
\cite{front23} are non-orthogonal, such  that special care has to be taken in terms of 
regularization of the equations of motion \cite{prb20}, however. This approach then led 
to an almost quantitative prediction also of the more elaborate quantum effects in 
the bosonic Josephson junction, using only a handful of basis functions, sometimes even 
as little as two. In addition, the fully variational approach has been shown
capable of converging to the full (Fock space) quantum results for the dynamics with
less computational effort and a possibly much better scaling than
the factorial scaling in terms of particle and site number of the Fock space
approach also for systems with more than just two wells \cite{pra21}. 

Restricting the Hamiltonian underlying the dynamics to be separable, we will show
in the following that exact results for the quantum dynamics can be gained, 
not only for Glauber CS but also for generalized coherent states. This is worthwhile 
from a mathematical standpoint but may also be helpful if benchmarks for 
non-separable programming and debugging tasks are asked for. Furthermore,
the occurrence of sharp structures in the so-called dynamical free energy
at certain points in time  and their dependence on parameters like the filling factor 
can be studied.

The presentation is organized as follows. In Section \ref{sec:GCS}, the definition
of the generalized coherent states that we will frequently use 
as initial states is reviewed and their relation to  the more commonly known Glauber 
or field coherent states is highlighted.
In Section \ref{sec:model}, this is followed by a review of results from the literature 
for the dynamics of interacting bosons in deep lattices. The focus in this section is 
on both, standard as well as generalized coherent state initial conditions and on the 
calculation of two-point correlation functions. The thermodynamic limit will then be 
considered, where both cases coincide. In Section \ref{sec:auto} we go beyond
charted territory by studying the relation between the (auto-)correlations and also 
the dynamical free-energy densities for the two different types of coherent states.
It will be shown that in the thermodynamic limit, the correlations are related by 
a Fourier type relation. This fact is finally shown to be true also in the case of 
a general Hamiltonian, beyond the separable deep lattice case. Along the way, we
give an explanation in terms of Husimi functions of time evolved Glauber CS
for the sharp structures in the dynamic free energy.
Finally, conclusions and an outlook are given and some analytical manipulations for the 
time evolution of a GCS are contained in the appendix.

\section{Definition of generalized coherent states}\label{sec:GCS}
 
In the following, the concept of a GCS will be reviewed along with a comprehensive 
discussion of its fundamental features, including a physical explanation of the state. 
Additionally, a key focus of this  section is to highlight the relationship between 
generalized coherent states and the more standard, so-called Glauber CS. 

As pointed out by Zhang et al \cite{ZFG90}, there exist several different kinds of 
GCS,  which can be obtained by generalizing the concept of Glauber CS via group 
theory. 
Herein, we focus on one class of them, which are also known as SU($M$) CS. The 
most widely used representation of the GCS, we employ is given by \cite{BP08,TWK09}:
\begin{eqnarray}
\label{wh}
|S,\vec{\xi}\rangle=\frac{1}{\sqrt{S!}}\left(\sum_{j=1}^M\xi_j\hat{a}_j^\dag\right)^S|0,0,\cdots,0\rangle,
\end{eqnarray}
where $\hat{a}_j^\dag$ represents the bosonic creation operator acting on the $j$-th mode or site in a lattice system and the bosonic 
commutator relation
\begin{equation}\label{eq:commutation}
    [\hat{a}_i,\hat{a}_j^\dagger]=\delta_{ij}
\end{equation}
is fulfilled by the ladder operators, which act on the number states
as follows
\begin{eqnarray}
\hat{a}|n\rangle&=\sqrt{n}|n-1\rangle,
\\
\hat{a}^\dagger|n\rangle&=\sqrt{n+1}|n+1\rangle.
\end{eqnarray}
Furthermore, the ket $|0,0,\cdots,0\rangle$ denotes the multi-mode vacuum state, and $S$ is the number of bosons in the GCS. The set of complex numbers $\{\xi_j\}$ are characteristic parameters of the GCS, satisfying the normalization condition
\begin{eqnarray}
\label{eq:norm}
\sum^{M}_{j=1}|\xi_j|^2=1,
\end{eqnarray}
where $M$ is the number of different modes. The physical interpretation of $|\xi_j|^2$ is the 
normalized population density or probability for the particles to be located on the $j$-th 
site. If, e.g., only a single $\xi_j$ is one and all the others are zero, this means that all
$S$ particles are located in the $j$th site.

Employing the general binomial (or multinomial) theorem, a GCS can be expanded in terms of 
the Fock states
\begin{eqnarray}\label{eq:GCS_Fock}
|S,\vec{\xi}\rangle=\sum_{[n_i]=S}\sqrt{S!}\prod_{i=1}^M\frac{1}{n_i!}(\xi_i\hat{a}_i^\dag)^{n_i}|0,0,\cdots,0\rangle,
\end{eqnarray}
where the sum $\sum_{[n_i]=S}$ is the short-hand notation of $\sum_{n_1+n_2+\cdots+n_M=S}$, 
which accounts for all possible configurations satisfying the constraint on the total particle 
number. The parameters $S$ and $M$ characterizing the system size determine the dimension of 
the Hilbert space
\begin{equation}
 C^{M-1}_{M+S-1}= \frac{(M+S-1)!}{S!(M-1)!},   
\end{equation}
spanned by the Fock states. This is the same as the number of possible 
configurations produced by placing $S$ identical balls into $M$ different boxes.

\subsection{Momentum space}

From the definition in Eq.\ (\ref{wh}), a GCS can be understood as a special form of 
condensate, where all particles macroscopically occupy a single-particle state given by 
$\sum_{j=1}^M \xi_j \hat{a}_j^\dag|0,0,\cdots,0\rangle$.  This non-local nature of the single-
particle state implies that any particle can be found at any site, with probabilities 
determined by $|\xi_i|^2$.\\
To further illustrate the concept of a condensate, we can represent the GCS in the quasi-
momentum frame. To this end, we consider a 1D lattice characterized by a lattice constant $a$. 
Next, we introduce a set of creation operators $\hat{b}_k^\dag$, which act on the quasi-
momentum $k$, and are related to the $\hat{a}_j^\dag$ through a Fourier transformation:
\begin{eqnarray}
\label{eq:bk}
    \hat{b}_k^\dag=\frac{1}{\sqrt{M}}\sum_{j=1}^M \hat{a}_j^\dag {\rm e}^{{\rm i}jka},
\end{eqnarray}
with $k$ ranging from $-\frac{\pi}{a}$ to $\frac{\pi}{a}$. Particularly 
noteworthy is the case of zero quasi-momentum, for which we have 
\begin{equation}
\label{eq:b=0}
 \hat{b}_{k=0}^\dag=\frac{1}{\sqrt{M}}\sum_{j=1}^M\hat{a}_j^\dag.   
\end{equation}
Homogeneous characteristic parameters of the GCS, i.e., 
$\xi_j=\frac{1}{\sqrt{M}}$ for all sites will cause all the particles 
to condense into the zero-momentum state
\begin{eqnarray}
    \frac{1}{\sqrt{S!}}\Big(\sum_{j=1}^M\frac{1}{\sqrt{M}}\hat{a}_j^\dag\Big)^S|0,0,\cdots,0\rangle=\frac{1}{\sqrt{S!}}(\hat{b}_{k=0}^\dag)^S|0,0,\cdots,0\rangle,
\end{eqnarray}
which is the so-called Bose-Einstein condensate \cite{Getal02}.

\subsection{Relation with multi-mode Glauber coherent states}

The GCS described above exhibit a close connection with the 
so-called Glauber CS. For a single harmonic mode,  whose Hamiltonian is given by
\begin{eqnarray}
\hat H=\frac{\hat p^2}{2}+\frac{\omega^2}{2}\hat x^2,
\end{eqnarray}
the Glauber CS is defined through the action of the displacement operator 
on the ground state \cite{Glauber}, and reads
\begin{eqnarray}\label{eq:single_CS}
    |\alpha\rangle\equiv{\rm e}^{-\frac{|\alpha|^2}{2}}{\rm e}^{\alpha \hat{a}^\dag}|0\rangle={\rm e}^{-\frac{|\alpha|^2}{2}}\sum_{n=0}^\infty\frac{\alpha^n}{\sqrt{n!}}|n\rangle.
\end{eqnarray}
As can be inferred from the definition above, a Glauber CS is an eigenstate of the 
annihilation operator
\begin{eqnarray}
    \hat{a}|\alpha\rangle=\alpha|\alpha\rangle,
\end{eqnarray}
where the eigenvalue $\alpha$ is a complex number
\begin{eqnarray}\label{eq:cs_pq}
    \alpha=\frac{\omega^{1/2}q+{\rm i}\omega^{-1/2}p}{\sqrt{2}}
\end{eqnarray}
in terms of the expectation values of position and momentum,
denoted by $q$, respectively $p$, and also the width parameter $\omega$, which determines the uncertainty of position and momentum. 
The wavefunction of the Glauber CS in position representation is 
given by a Gaussian times a plane wave, according to
\begin{eqnarray}
    \langle x|\alpha\rangle=\left(\frac{\omega}{\pi}\right)^{\frac{1}{4}}\exp\left[-\frac{\omega}{2}(x-q)^2+{\rm i}p(x-\frac{q}{2})\right],
\end{eqnarray}
where the Klauder phase convention has been used \cite{KlauSkag}. For a more recent review,
including further properties of Glauber CS and their use in quantum dynamics, we refer 
to \cite{irpc21}. 

To make contact with the GCS, the above concept has to be extended towards multi-mode 
Glauber CS (MMGS), represented as product states of individual single-mode Glauber CSs.
As demonstrated in \cite{BP08} and corroborated by the Taylor expansion of the exponential 
function, the MMGS denoted by 
$|\vec{\alpha}\rangle=\prod_{i=1}^M|\alpha_i\rangle$,
is linked to the GCS through the following expression
\void{
\begin{eqnarray}\label{eq:expansion}
|\vec{\alpha}\rangle\non\ &={\rm e}^{-\frac{1}{2}\sum_{i=1}^M|\alpha_i|^2}{\rm e}^{\sum_{i=1}^M\alpha_i\hat{a}_i^\dag}|0,0,\cdots,0\rangle\\
\non &={\rm e}^{-\frac{1}{2}\sum_{i=1}^M|\alpha_i|^2}\sum_{S=0}^{\infty}\frac{1}{S!}(\sum_{i=1}^M\alpha_i\hat{a}_i^\dag)^S|0,0,\cdots,0\rangle\\
\non\ &={\rm e}^{-\frac{\tilde{N}}{2}}\sum_{S=0}^{\infty}\frac{\tilde{N}^{\frac{S}{2}}}{\sqrt{S!}}\frac{1}{\sqrt{S!}}\Big(\sum_{i=1}^M{\frac{\alpha_i}{\sqrt{\tilde{N}}}}\hat{a}_i^\dag\Big)^S|0,0,\cdots,0\rangle\\
&={\rm e}^{-\frac{\tilde{N}}{2}}\sum_{S=0}^{\infty}\frac{\tilde{N}^{\frac{S}{2}}}{\sqrt{S!}}|S,\vec{\xi}\rangle.
\end{eqnarray}
}
\begin{equation}\label{eq:expansion}
|\vec{\alpha}\rangle\non\ =\sum_{S=0}^{\infty}\sqrt{P(S)}|S,\vec{\xi}\rangle,
\end{equation}
with
\begin{eqnarray}
P(S)= {\rm e}^{-{\tilde{N}}}\tilde{N}^{S}/{S!}. 
\end{eqnarray}
Here, $\tilde{N}=\sum_{i=1}^M|\alpha_i|^2$ denotes the average number of bosons in 
$|\vec{\alpha}\rangle$. We note that the relationship between $\xi_i$ and $\alpha_i$ is given 
by\[\xi_i=\frac{\alpha_i}{\sqrt{\tilde{N}}},\] which is reasonable, because $|\alpha_i|^2$ 
denotes the average number of particles while $|\xi_i|^2=\frac{|\alpha_i|^2}{\tilde{N}}$ 
corresponds to the normalized population density. 

Furthermore, we note that if all the $\alpha_i$ share the same phase, the 
corresponding GCS parameter is given by $\xi_i=\frac{|\alpha_i|}{\sqrt{\tilde{N}}}$, 
with a global ($S$-dependent) phase factor in front of every GCS in the sum in Eq.\ (\ref{eq:expansion}). Finally, the expression for the expansion coefficient $P(S)$ given above 
proves that the total number of particles follows a Poisson distribution with the 
mean $\tilde{N}$.
Eq.\ (\ref{eq:expansion}) establishes that the  MMGS is a superposition of GCS 
with varying particle numbers $S$, or equivalently speaking, one GCS with particle number $S$ 
can be acquired by projecting the MMGS onto the S-particle subspace. Consequently, it is 
natural to construct a one-to-one mapping between the GCS and MMGS.

The similarities and differences between the two sets of states are summarized in 
Table \ref{tab:comp}.
The definitions of MMGS and GCS are presented in 
the second line of the table, where it becomes evident that the 
Glauber CS is a 
product state, while the GCS cannot generally be factorized. In the third 
line, we examine the impact of annihilation operators on the two states. 
The Glauber CS emerges as an eigenstate of the annihilation operator 
$\hat{a}_i$ with the eigenvalue $\alpha_i$. In contrast, $\hat{a}_i$ 
annihilates a 
particle from the original GCS and generates a new GCS with the particle 
number $S-1$. Moving to the fourth line, the computation of the two-point 
correlation function is facilitated by the insights from the third line. 
The correlation functions of both states remain independent of the 
spatial separation between points $i$ and $j$, which proves that Glauber 
CS and GCS have long-range phase coherence. In the next line, we present the 
inner product between two MMGS and two GCS, respectively, both of 
which are non-orthogonal. The overlap of two different MMGSs has an 
exponential form, and for GCS, the overlap is expressed as a polynomial.
\renewcommand{\arraystretch}{2}
\begin{center}
\begin{table}
\begin{tabular}{c|c}

Multi-mode Glauber CS & GCS \\ 
\hline
$|\vec{\alpha}\rangle=\prod_{i=1}^M{\rm e}^{-\frac{|\alpha_i|^2}{2}}{\rm e}^{\alpha_i\hat{a}_i^\dag}|{\rm vac}\rangle=\otimes^M_{i=1}|\alpha_i\rangle$ & $|S,\vec{\xi}\rangle=\frac{1}{\sqrt{S!}}\left(\sum_{i=1}^M\xi_i\hat{a}_i^\dag\right)^S
|{\rm vac}\rangle$ \\ 
\hline
$\hat{a}_i|\vec{\alpha}\rangle=\alpha_i|\vec{\alpha}\rangle$ & $\hat{a}_i|S,\vec{\xi}\rangle=\sqrt{S}\xi_i|S-1,\vec{\xi}\rangle$\\ 
\hline
$\langle\vec{\alpha}|\hat{a}_i^\dag \hat{a}_j|\vec{\alpha}\rangle=\alpha_i^*\alpha_j$ & $\langle S,\vec{\xi}|\hat{a}_i^\dag \hat{a}_j|S,\vec{\xi}\rangle=S\xi_i^*\xi_j$ \\ 
\hline
$\langle\vec{\alpha}|\vec{\beta}\rangle=\exp\left[\sum_{i=1}^M\alpha^*_i\beta_i-\frac{1}{2}(|\alpha_i|^2+|\beta_i|^2)\right]$ & $\langle S',\vec{\xi}|S,\vec{\eta}\rangle=\left(\sum_{i=1}^M\xi_i^*\eta_i\right)^S\delta_{S,S'}$\\
\hline
overcomplete in whole Hilbert space & overcomplete in S-particle subspace
\end{tabular}
\caption{Comparison between MMGS and GCS}
\label{tab:comp}
\end{table}
\end{center}

\section{Model system and dynamics}\label{sec:model}

\subsection{Interacting-boson model}

To set the stage for the investigation of the quantum dynamics of a many-body 
system, we 
first review  some of the results of the seminal publication
by Schachenmayer  et al \cite{SDZ11}. To this end, 
we consider a nonlinear  Hamiltonian for interacting  bosonic quantum particles, 
distributed over  $M$ modes (wells) of the form
\begin{eqnarray}\label{hm}
 H=\frac{U}{2}\sum_{i=1}^M\hat{a}_i^{\dag 2}\hat{a}_i^2=\frac{U}{2}\sum_{i=1}^M\hat{n}_i(\hat{n}_i-1)=\sum_{i=1}^M\hat{h}_i,
\end{eqnarray}
where $\{\hat a_i,\hat a^\dagger_i\}$ are the creation and  annihilation  operators of
the previous section and $\hat n_i=\hat a_i^\dag\hat a_i$ is the corresponding number operator.
Furthermore,
\be
\hat{h}_i=\frac{U}{2}\hat{n}_i(\hat{n}_i-1)
\ee
stands for the local Kerr-type Hamiltonians with on-site
interaction strength $U$. The total Hamiltonian is the limiting case of 
vanishing hopping strength of the Bose-Hubbard model and can be realized in 
experiment by increasing the intensity of the optical lattice laser beams \cite{GMHB02}.

\subsection{Two-point correlation function for Glauber as well as generalized coherent states}

Other dynamical quantities that are frequently studied in the literature are 
the auto-correlation \cite{GMHB02,TSUUR18} as well as the two-point correlation function 
(TPCF) \cite{SDZ11}. In the following, we focus on the two-point correlation and
postpone the discussion of the auto-correlation to Sec.\ \ref{sec:auto}.

The analytical results to be derived for the cases of MMGS and GCS provide 
an insight for the connection between these two types of states. 
First, we revisit the results from \cite{SDZ11}, as their findings offer a comprehensive understanding of the relationship between the dynamics of Glauber as well as generalized
coherent states. We also  provide a more concise derivation for the case of GCS than the one in \cite{SDZ11}, by using some of the properties listed in Table.\ \ref{tab:comp}.

In the deep-lattice case, the two-point correlation function of a MMGS can be 
derived as follows \cite{SDZ11}
\begin{eqnarray}\label{eq:corrglauber}
 \langle\vec{\alpha}|{\rm e}^{{\rm i}\hat{H}t}\hat{a}^\dag_i\hat{a}_j{\rm e}^{-{\rm i}\hat{H}t}|\vec{\alpha}\rangle\non &=\langle\alpha_i|\langle\alpha_j|{\rm e}^{{\rm i}(\hat{h}_i+\hat{h}_j)t}\hat{a}_i^\dag \hat{a}_j{\rm e}^{-{\rm i}(\hat{h}_i+\hat{h}_j)t}|\alpha_i\rangle|\alpha_j\rangle\\
 \non &=\langle\alpha_i|{\rm e}^{{\rm i}\hat{h}_it}\hat{a}_i^\dag {\rm e}^{-{\rm i}\hat{h}_it}|\alpha_i\rangle\langle\alpha_j|{\rm e}^{{\rm i}\hat{h}_jt}\hat{a}_j{\rm e}^{-{\rm i}\hat{h}_jt}|\alpha_j\rangle\\
 \non &=\langle\alpha_i|\hat{a}_i^\dag {\rm e}^{{\rm i}U\hat{n}_it}|\alpha_i\rangle\langle\alpha_j|{\rm e}^{-{\rm i}U\hat{n}_jt}\hat{a}_j|\alpha_i\rangle\\
 \non &=\alpha_i^*\alpha_j\langle\alpha_i|\alpha_i{\rm e}^{{\rm i}Ut}\rangle\langle\alpha_j|\alpha_j{\rm e}^{-{\rm i}Ut}\rangle\\
 \non &=\alpha_i^*\alpha_j\exp\big[|\alpha_i|^2({\rm e}^{{\rm i}Ut}-1)\big]\exp\big[|\alpha_j|^2({\rm e}^{-{\rm i}Ut}-1)\big]\\
  &=|\alpha|^2{\rm e}^{2|\alpha|^2[\cos(Ut)-1)]}
\end{eqnarray}
where in the third line, we have used the formulas $\hat{a}_kf(\hat{n}_k)=f(\hat{n}_k+1)\hat{a}_k$ and $\hat{a}_k^\dag f(\hat{n}_k)=f(\hat{n}_k-1)\hat{a}_k^\dag$, and we obtain the result in the fourth line via the fact 
\begin{equation}
\label{eq:csphase}
{\rm e}^{{\rm i}\varphi \hat{n}_k}|\alpha_k\rangle=\sum_{n=0}^{\infty}\frac{\alpha_k^n}{\sqrt{n!}}{\rm e}^{{\rm i}\varphi n}|n\rangle_k=|\alpha_k {\rm e}^{{\rm i}\varphi}\rangle.   
\end{equation} 
For the last line, the system is assumed to be homogeneous: $\alpha_i=\alpha_j=\alpha$. From 
the final result, it is clear that the evolution of the correlation function is also 
driven by the on-site interaction effect, exhibiting a periodic behavior with a period 
of $2\pi/U$.

\begin{figure}[h]
\centering
\includegraphics[width=4in]{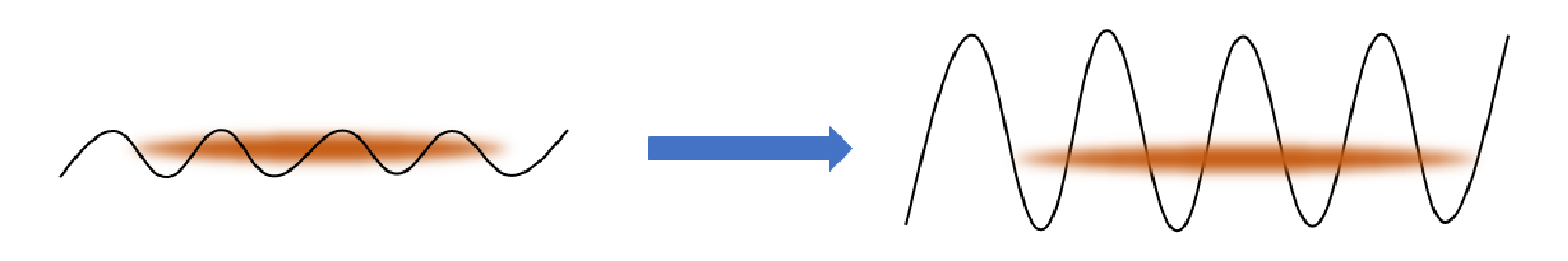}
\caption{Transition from the shallow lattice (almost free boson model) to deep lattice (on-site interaction only).}
\label{fig:lattice}
\end{figure}

As proposed in \cite{SDZ11}, we now choose the GCS defined in Eq.\ (\ref{wh}) as 
the initial state. Physically, this choice appears more reasonable compared to the 
Glauber CS, given the fact that the ground state of a definite number of particles 
for the nearest-neighbor hopping model with periodic 
boundary conditions
\begin{equation}
\label{eq:fb}
\hat{H}_{\rm NN}=-\sum_{\langle i,j\rangle}t_{i,j}\hat a_i^\dagger a_j    
\end{equation}
i.e., a Bose-Hubbard Hamiltonian, where only 
the hopping term with amplitudes $t_{ij}$ is present, is described by a GCS \cite{SDZ11,LDA22}. 
The particular setup to be studied in the following thus corresponds to a quench dynamics: 
the lattice is initially shallow where particles can move freely among different sites, and 
the system is prepared in the ground state. At $t=0$ the lattice is abruptly deepened as shown 
schematically in Fig.\ \ref{fig:lattice}, so the initial state which is no longer an 
eigenstate will start to evolve.
\\
The dynamics of the  two-point correlation function of an initial 
GCS governed by the interacting boson Hamiltonian of Eq.\ (\ref{hm}) is 
given as follows:
\begin{eqnarray}\label{corr1}
 \langle S,\vec{\xi}|{\rm e}^{{\rm i}\hat{H}t}\hat{a}_i^\dag \hat{a}_j{\rm e}^{-{\rm i}\hat{H}t}|S,\vec{\xi}\rangle\non&=\langle S,\vec{\xi}|{\rm e}^{{\rm i}(\hat{h}_i+\hat{h}_j)t}\hat{a}_i^\dag \hat{a}_j{\rm e}^{-{\rm i}(\hat{h}_i+\hat{h}_j)t}|S,\vec{\xi}\rangle\\
 \non&=\langle S,\vec{\xi}|\hat{a}_i^\dag {\rm e}^{{\rm i}U\hat{n}_it} {\rm e}^{-{\rm i}U\hat{n}_jt}\hat{a}_j|S,\vec{\xi}\rangle\\
 \non&=S\xi_i^*\xi_j\langle S-1,\vec{\xi}|{\rm e}^{{\rm i}U\hat{n}_it} {\rm e}^{-{\rm i}U\hat{n}_jt}|S-1,\vec{\xi}\rangle\\
 &=S\xi_i^*\xi_j\Big(|\xi_i|^2{\rm e}^{{\rm i}Ut}+|\xi_j|^2{\rm e}^{-{\rm i}Ut}+\sum_{k\neq i,j}^M|\xi_k|^2\Big)^{S-1},
\end{eqnarray}
where we have used the intermediate result
\begin{eqnarray}
 \hspace{-2.0cm}{\rm e}^{{\rm i}\varphi \hat{n}_k}|S,\vec{\xi}\rangle\non&={\rm e}^{{\rm i}\varphi \hat{n}_k}\frac{1}{\sqrt{S!}}\Big(\sum_{i=1}^M\xi_i\hat{a}_i^\dag\Big)^S|{\rm vac}\rangle\\
 \non&=\sum_{[n_i]=S}\frac{S!}{n_1!n_2!\cdots n_M!}{\rm e}^{{\rm i}\varphi \hat{n}_k}(\xi_1a_1^\dag)^{n_1}\cdots(\xi_ka_k^\dag)^{n_k}\cdots(\xi_M a_M^\dag)^{n_M}|
 {\rm vac}\rangle\\
 \non&=\sum_{[n_i]=S}\frac{S!}{n_1!n_2!\cdots n_M!}(\xi_1a_1^\dag)^{n_1}\cdots(\xi_ka_k^\dag)^{n_k}\cdots(\xi_M a_M^\dag)^{n_M}{\rm e}^{{\rm i}\varphi (\hat{n}_k+n_k)}|{\rm vac}\rangle\\
  \non&=\sum_{[n_i]=S}\frac{S!}{n_1!n_2!\cdots n_M!}(\xi_1\hat{a}_1^\dag)^{n_1}\cdots(\xi_k{\rm e}^{{\rm i}\varphi}\hat{a}_k^\dag)^{n_k}\cdots(\xi_M \hat{a}_M^\dag)^{n_M}{\rm e}^{{\rm i}\varphi (\hat{n}_k+n_k)}|{\rm vac}\rangle\\
  &=|S,\vec{\xi}^{'}\rangle,
\end{eqnarray}
where $\vec{\xi}^{'}=\{\xi_1,\xi_2,\cdots,\xi_k{\rm e}^{{\rm i}\varphi},\cdots,\xi_M\}$,  $\langle S,\vec{\xi}|S,\vec{\xi}^{'}\rangle=\big(|\xi_k|^2{\rm e}^{{\rm i}\varphi}+\sum_{i\neq k}^M|\xi_i|^2\big)^S$. This result illustrates that the effect of the number operator $\hat{n}_j$ is to imprint a phase factor on the $j$-th site, similar to the case of Glauber CS displayed in Eq.\ (\ref{eq:csphase}).  We note that the result in Eq.\ (\ref{corr1}) can 
also be obtained with the help of projection operator techniques \cite{SDZ11}.

For the special case where $i=j$, the correlation function simplifies to
\begin{eqnarray}
 \langle S,\vec{\xi}|{\rm e}^{{\rm i}\hat{H}t}\hat{a}_i^\dag \hat{a}_i{\rm e}^{-{\rm i}\hat{H}t}|S,\vec{\xi}\rangle\non&=\langle S,\vec{\xi}|{\rm e}^{{\rm i}\hat{h}_it}\hat{a}_i^\dag \hat{a}_i{\rm e}^{-{\rm i}\hat{h}_it}|S,\vec{\xi}\rangle\\
 \non&=\langle S,\vec{\xi}|\hat{a}_i^\dag \hat{a}_i|S,\vec{\xi}\rangle\\
 &=S|\xi_i|^2
\end{eqnarray}
For simplicity, we consider the homogeneous state $\xi_1=\xi_2=\cdots=\xi_M=\frac{1}{\sqrt{M}}$, which corresponds to the zero quasi-momentum state of Eq.\ (\ref{eq:b=0}). 
Under this condition, the correlation function Eq.\ (\ref{corr1}) simplifies to
\begin{eqnarray}
    \langle \hat{a}_i^\dag \hat{a}_j\rangle=\frac{S}{M}\left[1+\frac{2}{M}\cos(Ut)-\frac{2}{M}\right]^{S-1}
\end{eqnarray}\\
In the thermodynamic limit, $S\to \infty$, $M\to\infty$ with finite filling factor
\be
\lambda=\frac{S}{M},
\ee
Eq.\ (\ref{corr1}) converges to
\begin{eqnarray}\label{corr2}
    \lim_{S\to \infty}\langle \hat{a}_i^\dag \hat{a}_j\rangle&=\lim_{S\to \infty}\frac{S}{M}\left[1+\frac{S}{M}\frac{2}{S}\cos(Ut)-\frac{S}{M}\frac{2}{S}\right]^{S-1}\nonumber\\
    &=\lambda{\rm e}^{\lambda[-2+2\cos(Ut)]}.
\end{eqnarray}
 The similarity between Eq.\ (\ref{eq:corrglauber}) and Eq.\ (\ref{corr2}) reveals that the relationship between Glauber CS and GCS can be established by setting $\alpha=\sqrt\lambda$, which is also corroborated by Eq.\ (\ref{eq:expansion}). This demonstrates that the correlation function of GCS is identical to that of the Glauber CS in the thermodynamic limit. As proven in \cite{SDZ11},
 this identity is achieved already for rather small values of $S$.

\section{Auto-correlation function of the evolved GCS}\label{sec:auto}

Although GCS and Glauber CS share similarities for dynamical quantities like the two-point 
correlation functions and thus also the quasi-momentum distribution, differences between 
them still exist, even in the thermodynamic limit. To prove this fact, we will investigate 
the time-dependent auto-correlation function of the GCS in the following and express the result 
in terms of a related quantity for MMGS. 

 \subsection{Dynamical free energy density}
Given the similarity between the auto-correlation function 
\be
\langle\Psi(0)|{\rm e}^{-{\rm i}\hat{H}t}|\Psi(0)\rangle
\ee
and the partition function in equilibrium statistical mechanics
\begin{eqnarray}
{\rm Z}={\rm tr}({\rm e}^{-\beta\hat{H}})={\rm e}^{-\beta Nf},
\end{eqnarray}
where $\beta=1/kT$ is proportional to the inverse of temperature, $N$ denotes the number of 
degrees of freedom and $f$ is the free-energy density, an analog of thermodynamic
phase transitions, i.e., a nonanalytic behaviour at critical times in  quantum dynamics 
was predicted \cite{HPK13,KKK14,FUBSG17,Heyl18}.

In the following, we consider the dynamical free energy, defined by
\begin{eqnarray}\label{eq:echo}
    \mathcal{L}\equiv-\frac{1}{M}\log(|\langle\Psi(0)|\Psi(t)\rangle|^2),
\end{eqnarray}
where the initial state in our case
\begin{equation}
  |\Psi(0)\rangle=|S,\vec{\xi}\rangle 
\end{equation}
is the GCS with homogeneous parameters. As proven in the appendix, 
the auto-correlation function (or Loschmidt amplitude) is then given by
\begin{eqnarray}\label{autoc}
\langle\Psi(0)|\Psi(t)\rangle=
 \langle S,\vec{\xi}|{\rm e}^{-{\rm i}\hat{H}t}|S,\vec{\xi}\rangle=S!\sum_{[n_i]=S}\prod_{i=1}^M\frac{|\xi_i|^{2n_i}}{n_i!}{\rm e}^{-{\rm i}\frac{U}{2}n_i^2t}
\end{eqnarray}
and thus the dynamical free 
energy density \cite{LaHe19} is, at least in principle, at our disposal. 

Although the summation in Eq.\ (\ref{autoc}) has no obvious closed solution for 
$U\neq 0$, the numerical solution is accessible, however, via the concept of generating 
functions  \cite{Wilf05,MastLind} that allows for a relatively straightforward evaluation of the 
restricted  summation in Eq.\ (\ref{autoc}), which would otherwise be impossible to tackle, 
already for moderate particle and site numbers. 
The general idea of the generating function method is to design a polynomial in the variable  
$x$ with the constraint that the coefficient for the term $x^S$ is related to 
Eq.\ (\ref{autoc}). This method turns the problem into calculating the 
polynomial coefficients which can be done by efficient convolution algorithm. 
For instance, if the generating function is the product of two polynomials, represented as 
$\left(\sum_{i=1}^{\infty}a_ix^i\right)\left(\sum_{j=1}^{\infty}b_jx^j\right)$, the 
coefficient in front of the $x^{n}$ can be expressed as 
\[
\sum_{i=1}^\infty a_ib_{n-i},
\]
which is the convolution between two series with coefficients $\{a_i\}$ and $\{b_i\}$.

Let us first consider the  homogeneous situation 
$\xi_1=\xi_2=\cdots=\xi_M=\frac{1}{\sqrt{M}}$, simplifying Eq.\ (\ref{autoc}) to
\begin{eqnarray}\label{autoc2}
\langle S,\vec{\xi}|{\rm e}^{-{\rm i}\hat{H}t}|S,\vec{\xi}\rangle=\frac{S!}
{M^S}\sum_{[n_i]=S}\prod_{i=1}^M\frac{1}{n_i!}{\rm e}^{-{\rm i}\frac{U}{2}n_i^2t}
\end{eqnarray}
We define a polynomial 
\begin{eqnarray}
F(x)=\left[\sum_{k=0}^{M}\phi(k)x^{k}\right]^M,\indent \phi(k)=\frac{{\rm e}^{-{\rm i}\frac{U}{2}k^2t}}{k!}
\end{eqnarray}
which plays the role of generating function, so the result in Eq.\ (\ref{autoc2}) is given by $\frac{S!}{M^S}[x^S]F(x)$, where $[x^S]F(x)$ denotes the coefficient for term $x^S$ in the polynomial $F(x)$. In the general situation, a series of independent polynomials are needed
\begin{eqnarray}
\tilde{F}(x)=f_1(x)f_2(x)\cdots f_M(x),\indent f_i(x)=\sum_{k=0}^M\frac{|\xi_i|^{2k}{\rm e}^{-{\rm i}\frac{U}{2}k^2t}}{k!}x^k
\end{eqnarray}
and Eq.\ (\ref{autoc}) is given by $S! [x^S]\tilde{F}(x)$.

\begin{figure}[h]
\centering
\includegraphics[width=4in]{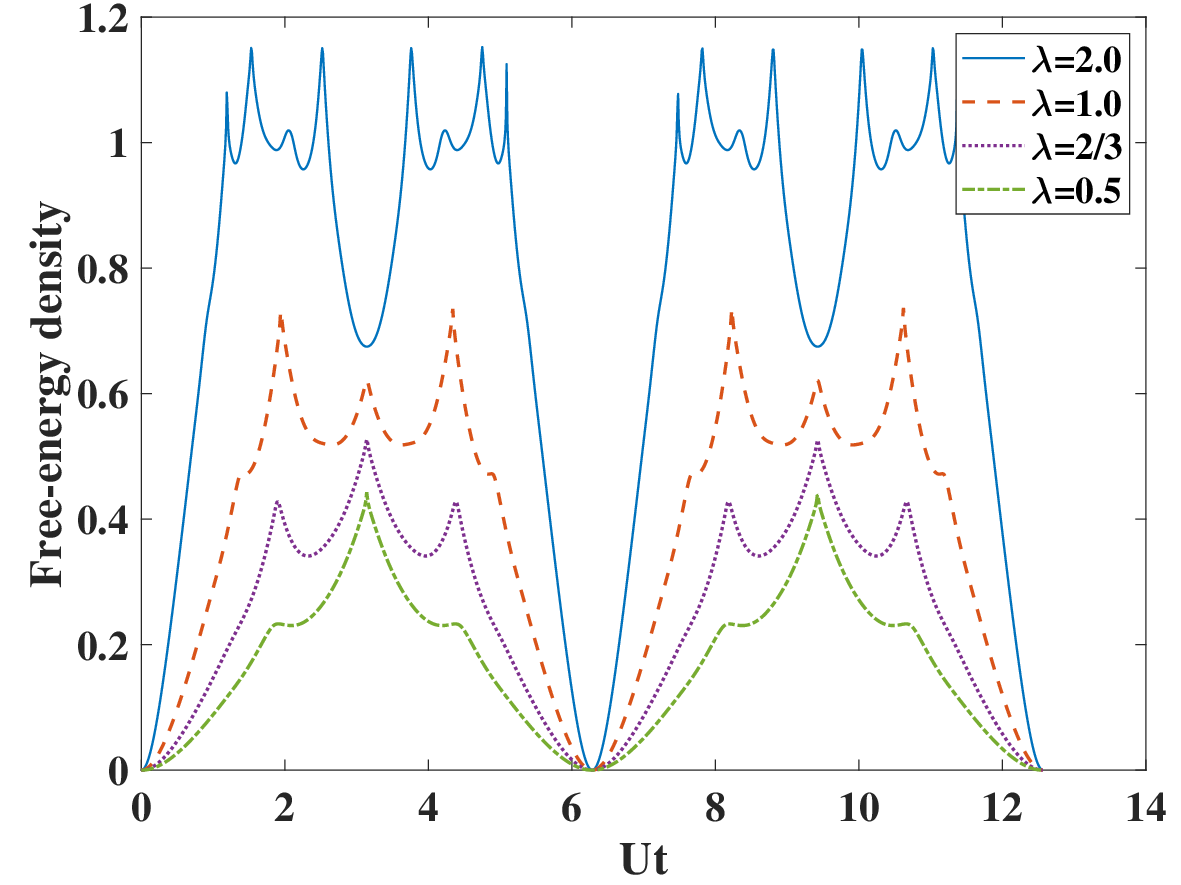}
\caption{Dynamical free-energy density for GCS initial condition as a function of time $Ut$, 
ranging from $0$ to $4\pi$.  The total number of particles is $S=100$ and results are 
shown for different values of filling factor $\lambda=S/M$}.
\label{fig:loschmidt}
\end{figure}

For the plot in Fig.\ \ref{fig:loschmidt}, we fix the value of 
the particle number to $S=100$ and vary the ratio $\lambda=\frac{S}{M}$ to investigate the dynamics. From Fig.\ \ref{fig:loschmidt}, we can observe that for large values of $\lambda$, the dynamical free-energy density has several peaks within one period. As $\lambda$ is decreased, these peaks gradually disappear, leaving just one peak occurring at $Ut=\pi$ when $\lambda=0.5$. Due to the definition of the dynamical free-energy density in Eq.\ (\ref{eq:echo}), these peaks correspond to the local minima
of the survival probability.
When the survival probability completely vanishes, non-analytical kinks appear in the curves. This phenomenon is recognized as a quantum dynamical phase transition (QDPT) in the literature \cite{Jurcevic17,zhang17,flaschner18}. The authors in \cite{LaHe19} 
have studied the QDPT of the non-equilibrium dynamics induced by the quench from the superfluid phase to the Mott-insulator phase present in the full Bose-Hubbard model, whereas
in \cite{LiSt20}, kink-like features in the rate function have been shown to exist in
a dissipative model of quantum optics (Dicke model) already at moderate particle number.

\begin{figure}[h]
\centering
\includegraphics[width=4in]{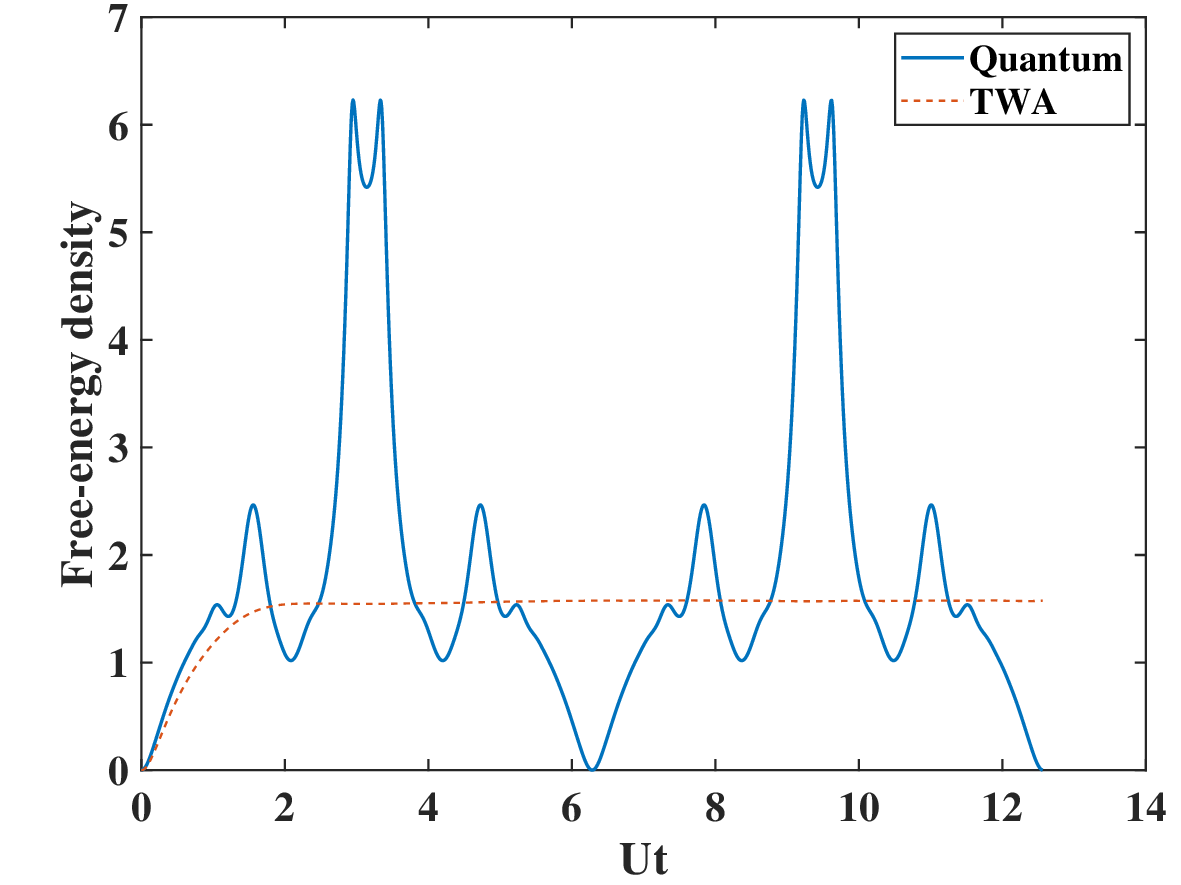}
\caption{Dynamical free-energy density for a Glauber state initial condition as a function of 
time $Ut$, ranging from $0$ to $4\pi$. The filling factor is $\lambda=2$. Solid line: 
full quantum result; dashed line: TWA result}.
\label{fig:loschmidt_2}
\end{figure}

In order to gain some understanding of the peak structure seen in 
Fig.\ \ref{fig:loschmidt}, we will now elucidate the connection of the 
auto-correlation with the one in the MMGS case, given by 
\begin{eqnarray}\label{eq:sur_prob_cs}
    \langle\vec{\alpha}|{\rm e}^{-{\rm i}\hat{H}t}|\vec{\alpha}\rangle\nonumber&=\otimes_{i=1}^M\langle\alpha_i|\exp(-{\rm i}\hat{h}_it)|\alpha_i\rangle\nonumber\\
    &=\prod_{i=1}^M {\rm e}^{-|\alpha_i|^2}\sum_{n_i=0}^{\infty}\frac{|\alpha_i|^{2n_i}}{n_i!}{\rm e}^{-{\rm i}\frac{U}{2}n_i(n_i-1)t}.
\end{eqnarray}
This result is factorized and does not impose constraints on the total number of particles
and is thus very easily accessible to numerical treatment. The corresponding 
free-energy density for the case of filling factor $\lambda=2$ is depicted by the solid 
line in Fig. \ref{fig:loschmidt_2} and shows a similar peak structure with minima
at $Ut$ equal to integer multiples of $2\pi$ as above. The corresponding 
classical result (starting from the same quantum mechanical initial condition),
calculated using the truncated Wigner approximation is comparatively structureless,
as displayed by the dashed line. The peaks can thus be considered as quantum effects.

To make use of the simple form of the auto-correlation in the Glauber case, we are
looking for a relation between the two cases. To find this relation, we introduce 
an irrelevant phase and rewrite the homogeneous GCS auto-correlation given 
in  Eq.\ (\ref{autoc}) as follows:
\begin{eqnarray}\label{eq:kronec}
\hspace{-2cm}
    \langle S,\vec{\xi}|{\rm e}^{-{\rm i}\hat{H}t}|S,\vec{\xi}\rangle&=S!\sum_{[n_i]=S}\prod_{i=1}^M\frac{1}{n_i!}\frac{1}{M^{n_i}}{\rm e}^{-{\rm i}\frac{U}{2}n_i(n_i-1)t}\nonumber\\
    &=S!\sum_{n_1=0}^{\infty}\sum_{n_2=0}^{\infty}\cdots\sum_{n_M=0}^{\infty}\delta_{S,\sum_{i=1}^Mn_i}\prod_{i=1}^M\frac{1}{n_i!}\frac{1}{M^{n_i}}{\rm e}^{-{\rm i}\frac{U}{2}n_i(n_i-1)t}\nonumber\\
    &=S!\sum_{n_1=0}^{\infty}\sum_{n_2=0}^{\infty}\cdots\sum_{n_M=0}^{\infty}\int_0^1{\rm d}x {\rm e}^{-{\rm i}2\pi x(S-\sum_{i=1}^Mn_i)}\prod_{i=1}^M\frac{1}{n_i!}\frac{1}{M^{n_i}}{\rm e}^{-{\rm i}\frac{U}{2}n_i(n_i-1)t}\nonumber\\
    &=S!\sum_{n_1=0}^{\infty}\sum_{n_2=0}^{\infty}\cdots\sum_{n_M=0}^{\infty}\int_0^1{\rm d}x {\rm e}^{-{\rm i}2\pi xS}\prod_{i=1}^M\frac{1}{n_i!}\frac{1}{M^{n_i}}{\rm e}^{-{\rm i}\frac{U}{2}n_i(n_i-1)t+{\rm i}2\pi xn_i}\nonumber\\
    &=S!\int_0^1{\rm d}x {\rm e}^{-{\rm i}2\pi xS}\left(\sum_{n=0}^{\infty}\frac{1}{n!}\frac{1}{M^{n}}{\rm e}^{-{\rm i}\frac{U}{2}n(n-1)t+{\rm i}2\pi  xn}\right)^M\nonumber\\
    &=\frac{S!}{S^S}\int_0^1{\rm d}x {\rm e}^{-{\rm i}2\pi xS}\left(\sum_{n=0}^{\infty}\frac{1}{n!}\frac{S^n}{M^{n}}{\rm e}^{-{\rm i}\frac{U}{2}n(n-1)t+{\rm i}2\pi  xn}\right)^M,
\end{eqnarray}
where in the first line, the global phase factor ${\rm e}^{{\rm i}\frac{U}{2}St}={\rm e}^{{\rm i}\frac{U}{2}\sum_{i=1}^Mn_it}$ is incorporated, in order to ease the comparison with the Glauber CS case. In the second line the Kronecker symbol $\delta_{S,\sum_{i=1}^Mn_i}$ is introduced to impose the constraint on the total particle number and meanwhile lift the restriction on the particle number of the single mode. The third line follows by replacing the Kronecker symbol with an integral
\begin{eqnarray}
    \delta_{S,\sum_{i=1}^Mn_i}=\int_0^1{\rm d}x {\rm e}^{-{\rm i}2\pi x(S-\sum_{i=1}^Mn_i)}.
\end{eqnarray}\\
Furthermore, by incorporating the factor $\frac{S!}{S^S}$ into the bracket, we are able to define the  function 
\begin{eqnarray}\label{eq:integrand}
    G(x,t)=\left[\sum_{n=0}^{\infty}\frac{1}{n!}\Big(\frac{S!}{S^S}\Big)^{1/M}\frac{S^n}{M^{n}}{\rm e}^{-{\rm i}\frac{U}{2}n(n-1)t+{\rm i}2\pi  xn}\right]^M
\end{eqnarray}
that will play a central role. 

Further progress can now only be made in the limit of large particle numbers, because in this 
case, we are able to use Stirling's approximation $S!\approx \sqrt{2\pi S}(\frac{S}{\rm e})^S$ 
of the factorial to simplify the factor $\Big(\frac{S!}{S^S}\Big)^{1/M}$ as
\begin{eqnarray}\label{eq:stirling}
    \Big(\frac{S!}{S^S}\Big)^{1/M}&\approx\sqrt[2M]{2\pi S}\left(\frac{S}{S{\rm e}}\right)^{\frac{S}{M}},
\end{eqnarray}
Combining Eq.\ (\ref{eq:kronec}), Eq.\ (\ref{eq:integrand}) and Eq.\ (\ref{eq:stirling}), we have
\begin{eqnarray}\label{eq:fourier}
    \langle S,\vec{\xi}|{\rm e}^{-{\rm i}\hat{H}t}|S,\vec{\xi}\rangle=\sqrt{2\pi S}\int_0^1{\rm d}x {\rm e}^{-{\rm i}2\pi xS}G(x,t)
\end{eqnarray}
where the $G(x,t)$ is expressed alternatively as
\begin{eqnarray}\label{eq:integrand2}
    G(x,t)=\left[\sum_{n=0}^{\infty}\frac{\lambda^n{\rm e}^{-\lambda}}{n!}{\rm e}^{-{\rm i}\frac{U}{2}n(n-1)t+{\rm i}2\pi  xn}\right]^M.
\end{eqnarray}
Inspired by the form of Eq.\ (\ref{eq:fourier}), we interpret the survival probability amplitude of the time-evolved GCS as the Fourier coefficient of $G(x,t)$ at a specific "frequency" denoted by $S$ \footnote{We note that the Fourier-transform pair of variables in the present case are $x$ and $S$}. 

By comparing Eqs.\ (\ref{eq:integrand2}) and (\ref{eq:sur_prob_cs}), we observe that
the function $G(x,t)$ is the cross-correlation of two different MMGS, i.e., 
\begin{eqnarray}
  G(x,t)=\langle \vec{\alpha'}|{\rm e}^{-{\rm i}\hat{H}t}|\vec{\alpha}\rangle,
\end{eqnarray}
where $|\vec{\alpha}\rangle$   
is a multimode Glauber CS with $\alpha_i=\sqrt{\lambda}$ for $\forall i$, and 
$|\vec{\alpha'}\rangle$ is a similar CS with a phase shifted parameter, characterized by  
\begin{equation}
 \alpha'_j=\sqrt{\lambda}{\rm e}^{-{\rm i}2\pi x}   
\end{equation} 
for $\forall j$ with 
\begin{equation}
   0\leq x<1.
\end{equation}
Despite the phase shift, $|\vec{\alpha}\rangle$ and $|\vec{\alpha'}\rangle$ are 
expanded by the  same GCS as discussed below Eq.\ (\ref{eq:expansion}). 
Essentially, this phase shift is rooted in the $U(1)$ symmetry (conservation of 
particle number) of the GCS, which the Glauber CS do not obey. Another significant 
insight gained from Eq.\ (\ref{eq:fourier}) is that if the 
value of $G(x)$ is zero, the survival probability will consequently be zero as well.

Before we continue the discussion of the dynamical free-energy density a few remarks
are in order: The generating function approach helps to get a numerical handle on 
the calculation of the GCS auto-correlation in the case of the Bose-Hubbard model 
with onsite interaction only. It will, however, not be generalizable to the 
dynamics including inter-site hopping. In contrast, the Fourier relation between 
the GCS auto-correlation and the cross-correlation of the Glauber coherent state 
(which is just a direct product in the deep lattice case), although it is 
derived for large particle numbers, is more general in terms of the Hamiltonians, 
it can be used for, as will be further elucidated below.

\subsubsection{Thermodynamic limit}

To be specific, we consider the case of unit filling factor $\lambda=1$, which 
corresponds to the thermodynamic limit when both $S$ and $M$ are large and equal. 
In this case, the function\footnote{This function is similar to the (finite) Jacobi 
theta sums studied in \cite{jpa97} in the context of quantum carpets but is more 
complex due to the presence of the factorial term as well as the infinite summation}
\begin{eqnarray}\label{eq:G_x}
    G(x,t)=\left[\sum_{n=0}^{\infty}\frac{{\rm e}^{-1}}{n!}{\rm e}^{-{\rm i}\frac{U}{2}n(n-1)t+{\rm i}2\pi  xn}\right]^M
\end{eqnarray}
represents the overlap function of the time-evolved MMGS 
${\rm e}^{-{\rm i}\hat{H}t}|\vec{\alpha}\rangle$ with MMGS distributed along the  
unit circle in phase space. 
\begin{figure}[h]
  \centering
  \includegraphics[width=2.5in]{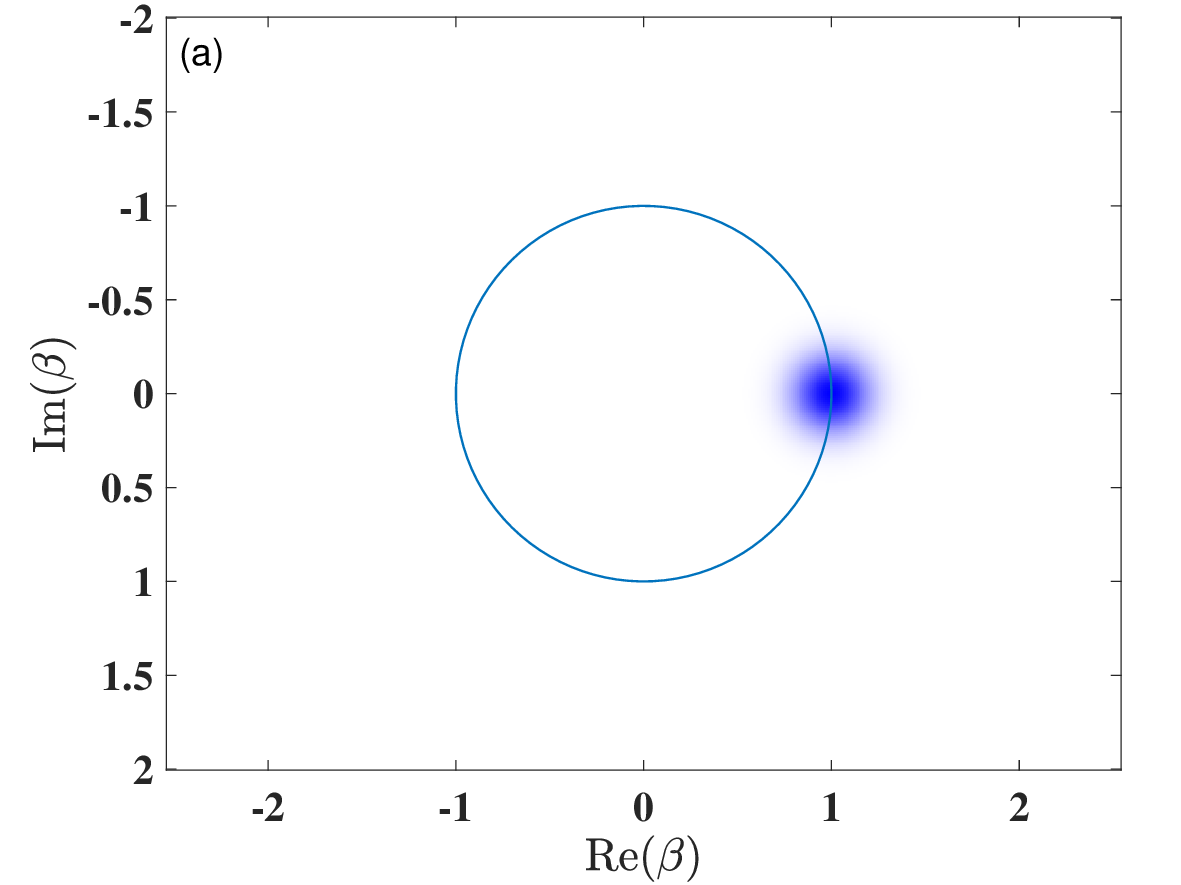}
  \includegraphics[width=2.5in]{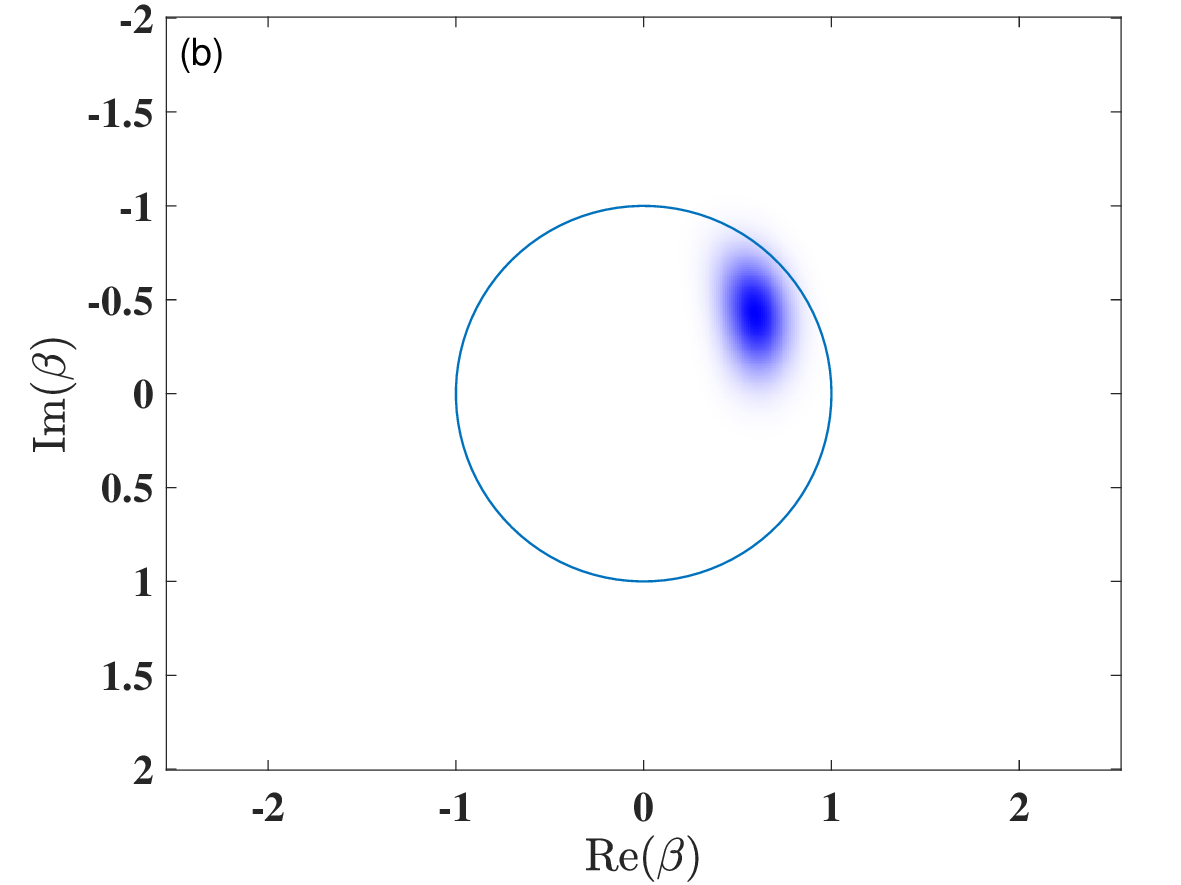}
  \includegraphics[width=2.5in]{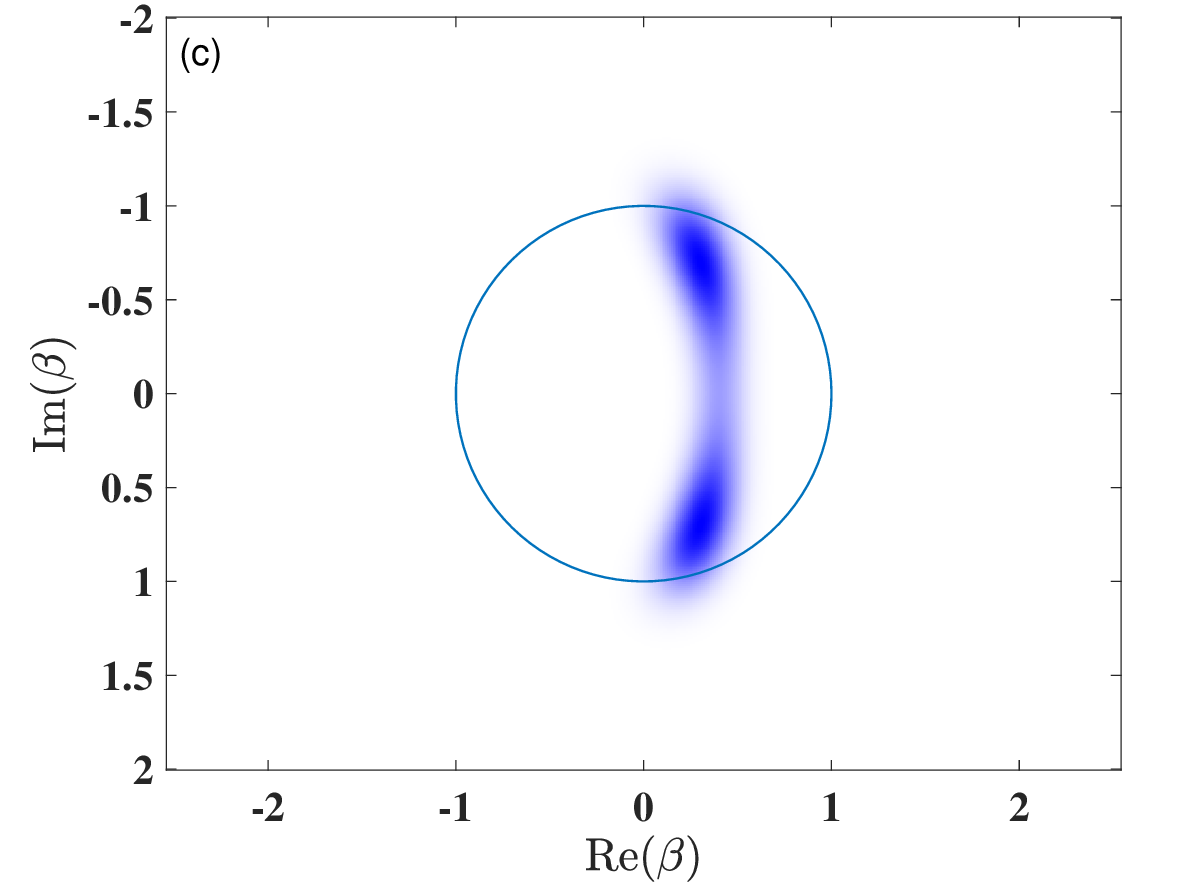}
  \includegraphics[width=2.5in]{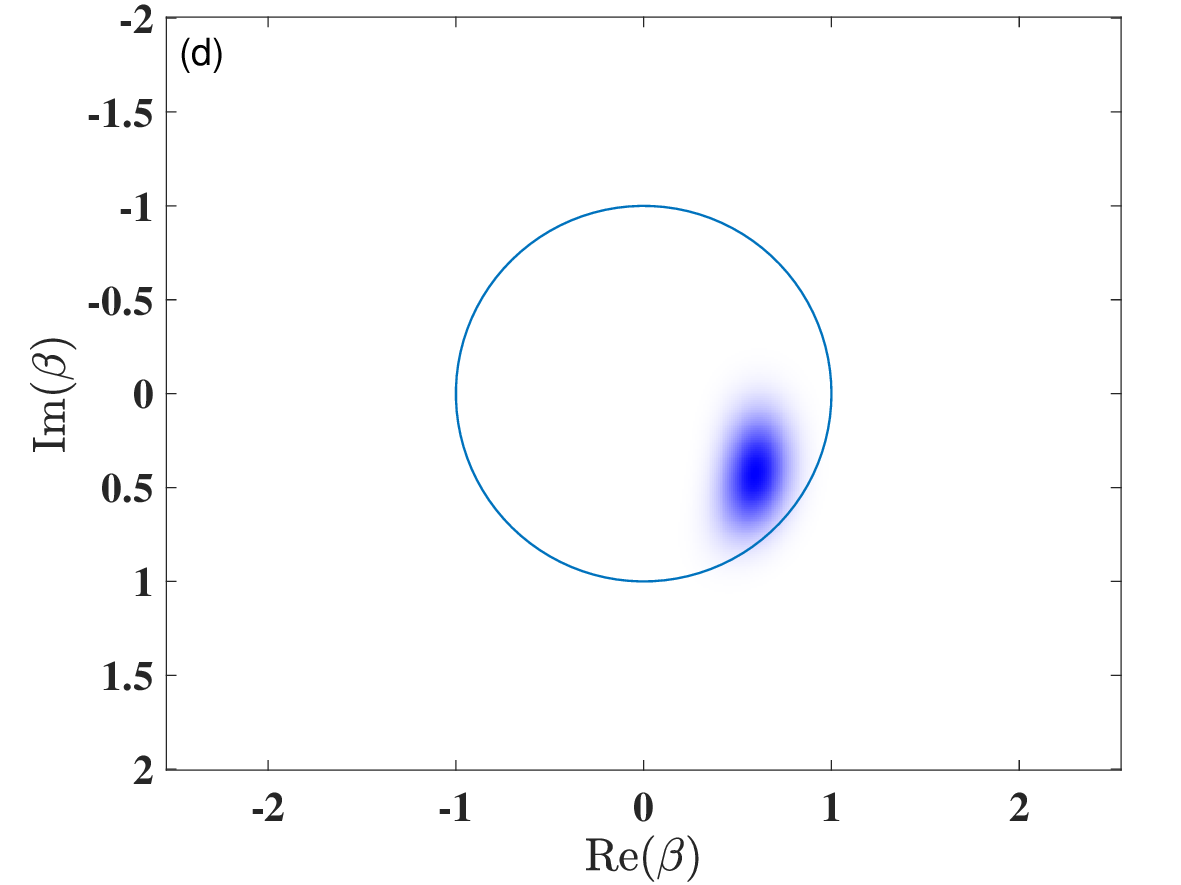}
  \\
  \caption{The distribution function $|\langle \beta|{\rm e}^{-{\rm i}\hat{H}t}|\alpha\rangle|^M$ in phase space $({\rm Re}(\beta),{\rm Im}(\beta))$ at different points in time: (a) $Ut=0$, (b) $Ut=\frac{\pi}{2}$, (c) $Ut=\pi$ and (d) $Ut=\frac{3\pi}{2}$. The characteristic CS parameter is $\alpha=1$ and the system parameter is $M=50$.}\label{fig:circle}
\end{figure}\\
In Fig.\ \ref{fig:circle}, we present the dynamics of the function 
$|\langle \beta|{\rm e}^{-{\rm i}\hat{H}t}|\alpha\rangle|^M$ 
with the single site cross-correlation
\begin{equation}
 c_{\beta\alpha}(t)\equiv\langle \beta|{\rm e}^{-{\rm i}\hat{H}t}|\alpha\rangle=
 \exp\left[-\frac{|\alpha|^2+|\beta|^2}{2}\right]\sum_{n=0}^\infty
 \frac{(\alpha\beta^\ast)^n}{n!}{\rm e}^{-{\rm i}\frac{U}{2}n(n-1)t}
 \label{eq:cc}
\end{equation}
for system size $S=M=50$. The value of $|G(x,t)|$ is determined by the 
distribution of $|\langle \beta|{\rm e}^{-{\rm i}\hat{H}t}|\alpha\rangle|^M$ 
on the unit circle. It can be inferred that this function
evolves around the origin, and the value of $|G(x,t)|$ is significantly influenced by 
the spot's relative position to the circle. Pictorially, when the circle does 
not intersect the center of the spot, the survival probability decreases, leading 
to the 
emergence of sharp structures in the rate function at these points in time. 
Moreover, as $M$ approaches infinity, the shape of the spot shrinks into a point, 
indicating that once the center of the point leaves the circle, the survival  
probability rapidly drops towards zero. Furthermore, we stress that in the 
present case,
the distribution function displayed in Fig.\ \ref{fig:circle}, at no instance in 
time is attaining appreciable weight for negative real part of $\beta$
in contrast to the case of larger values of $\lambda$ to be discussed below.

\begin{figure}[h]
\centering
\includegraphics[width=4in]{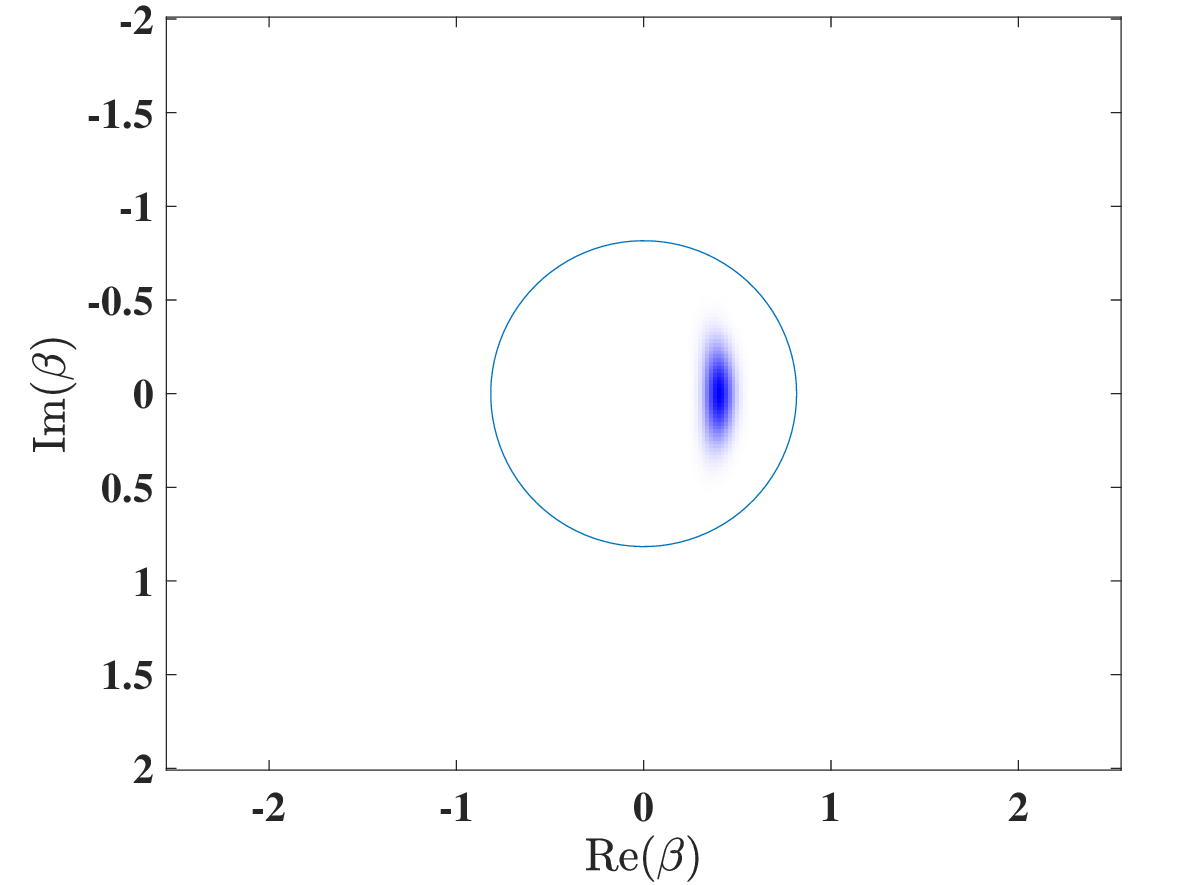}
\caption{The distribution function $|\langle \beta|{\rm e}^{-{\rm i}\hat{H}t}|\alpha\rangle|^M$ for $\lambda=\frac{2}{3}$ with $S=100$ at $Ut=\pi$.}
\label{fig:circle_pi}
\end{figure}

Finally, let us study the influence of the shape of the spot.  For example, when $\lambda$ is large the spot tends to split into two parts when $Ut=\pi$, possibly introducing extra intersections with the circle, as displayed in  Fig.\ \ref{fig:circle}\ (c).
Conversely, if $\lambda$ is small such as the case $\lambda=\frac{2}{3}$, the spot will maintain its shape and the center remains distanced from the circle at $Ut=\pi$ displayed in Fig.\ \ref{fig:circle_pi}, resulting in the gradual dominance of the central peak at $Ut=\pi$ for decreasing $\lambda$ as shown in Fig.\ \ref{fig:loschmidt}.\\

\subsubsection{The case of large filling factor}

\begin{figure}[h]
\centering
\includegraphics[width=4in]{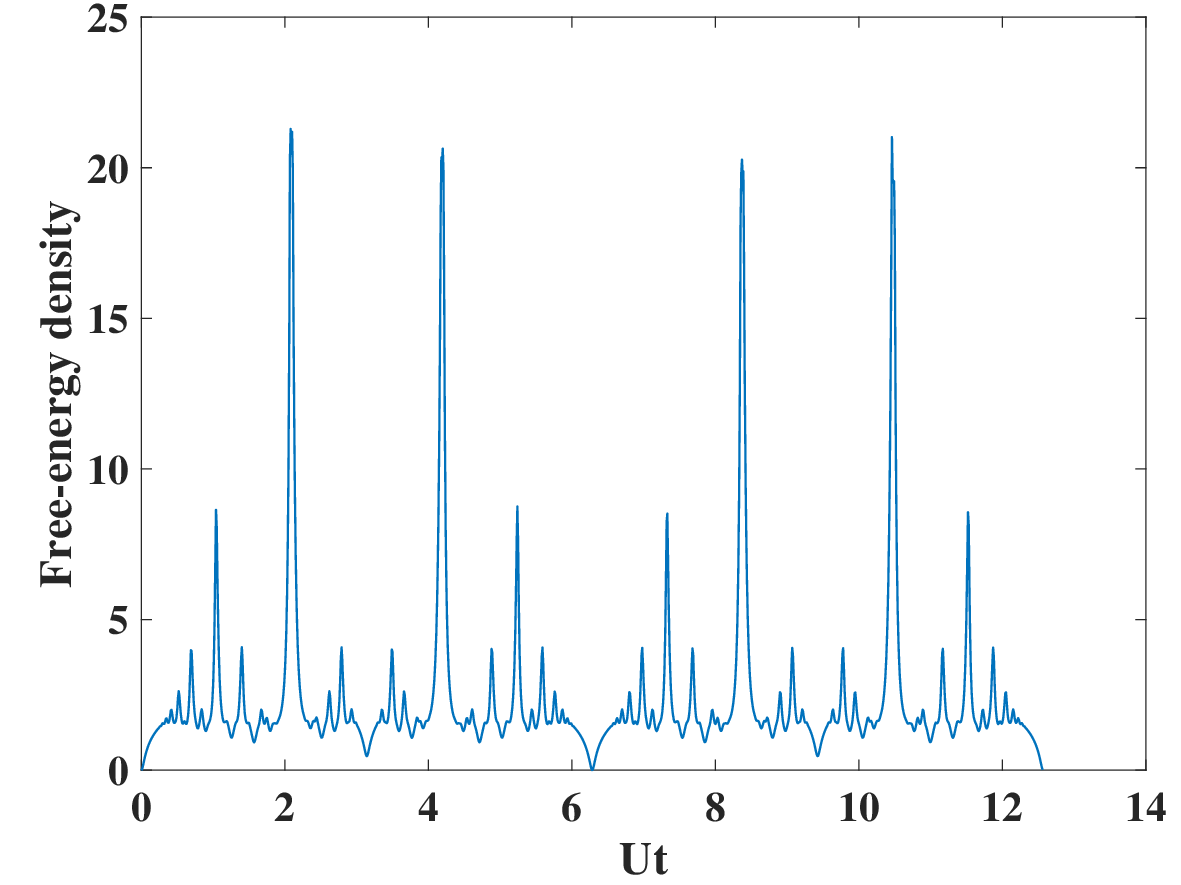}
\caption{Dynamical free-energy density 
as a function of time $Ut$ ranging from $0$ to $4\pi$. The system size is set to be $S=100$ and $M=3$.}
\label{fig:echo2}
\end{figure}

Another noteworthy scenario occurs when $S\gg M$, where the Eq.\ (\ref{eq:fourier}) 
still holds. However, the underlying distribution in phase space and the 
corresponding free-energy density are much more complicated. In contrast to the 
results in Fig.\ \ref{fig:loschmidt}, the free-energy density in Fig.\ 
\ref{fig:echo2} for $S=100$ and $M=3$ contains more peaks, especially two 
pronounced spikes located around $x=2.074$ and $x=4.2$ within the first period. 
These peaks are associated with the local or global minimal values of the survival 
probability.

\begin{figure}[htp]
\includegraphics[width=.3\textwidth]{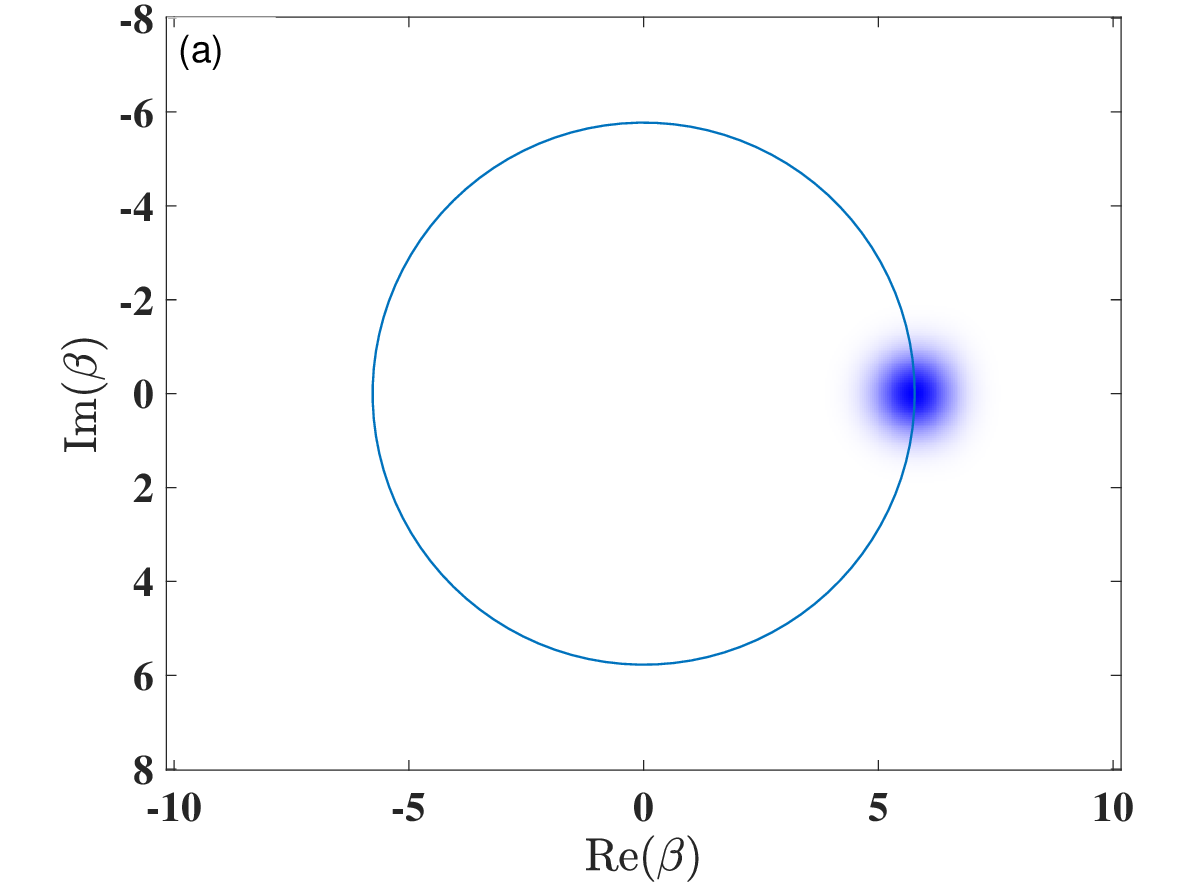}
\includegraphics[width=.3\textwidth]{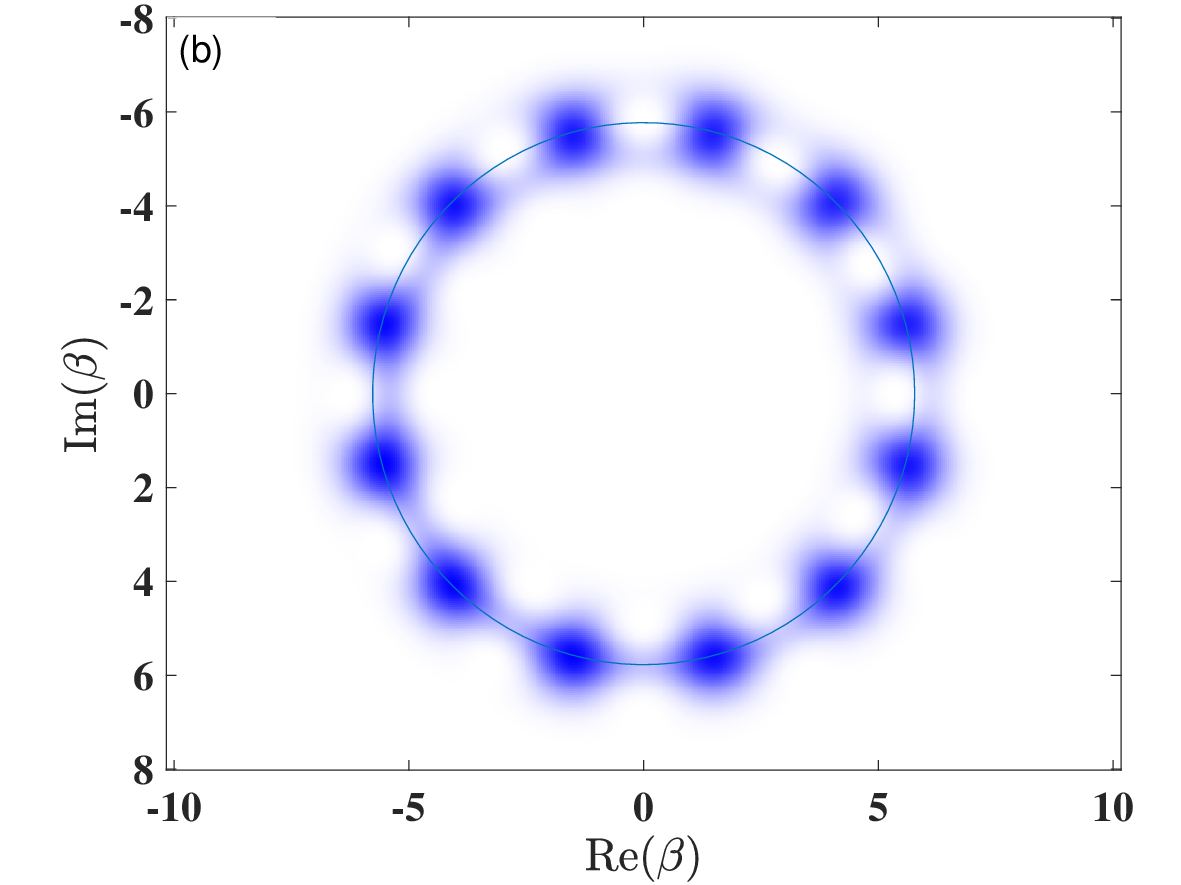}
\includegraphics[width=.3\textwidth]{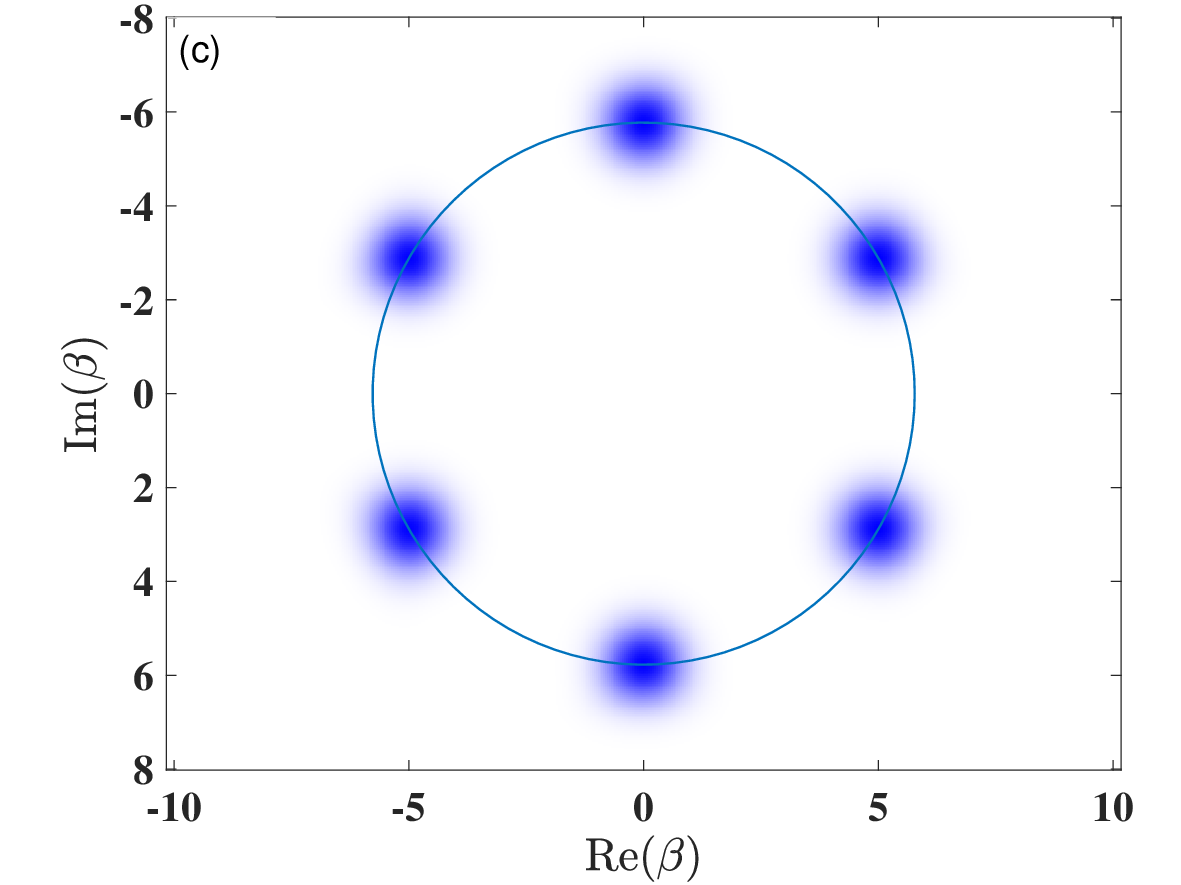}

\includegraphics[width=.3\textwidth]{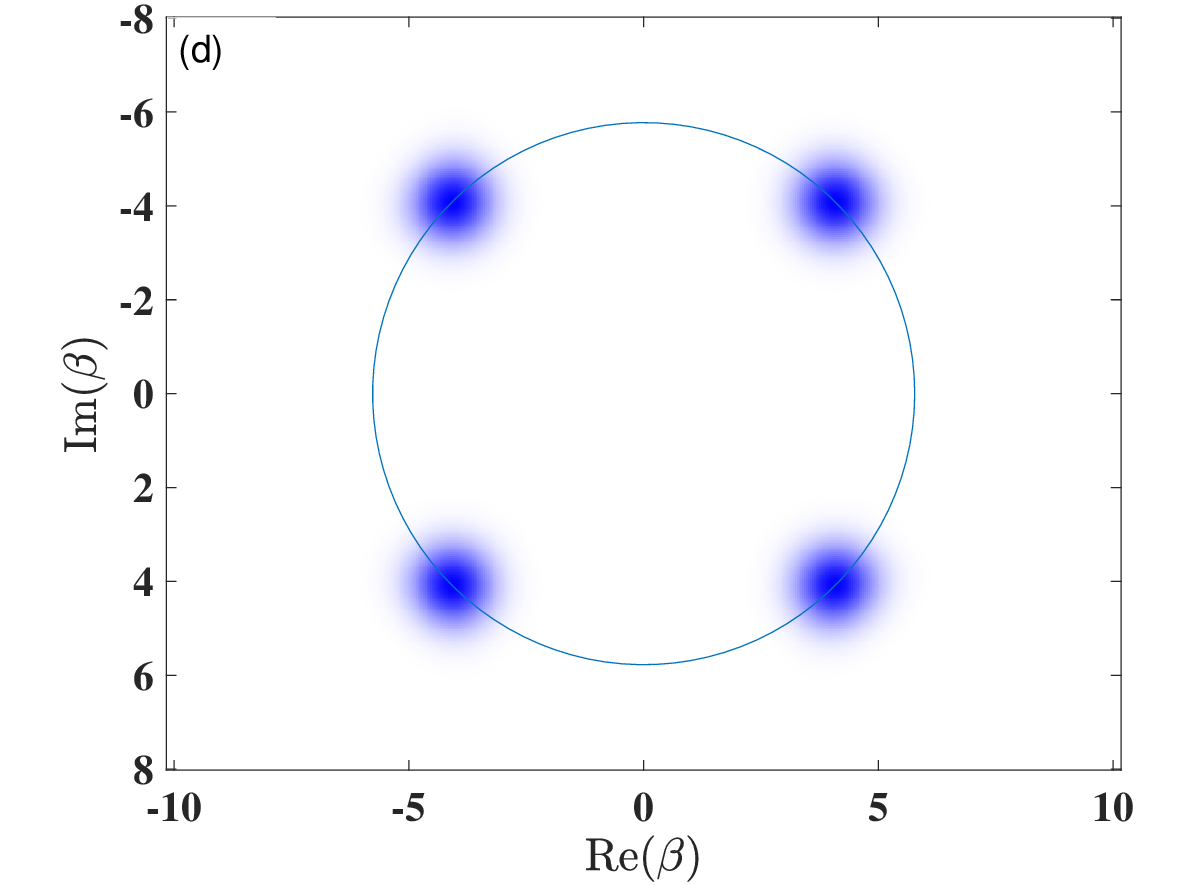}
\includegraphics[width=.3\textwidth]{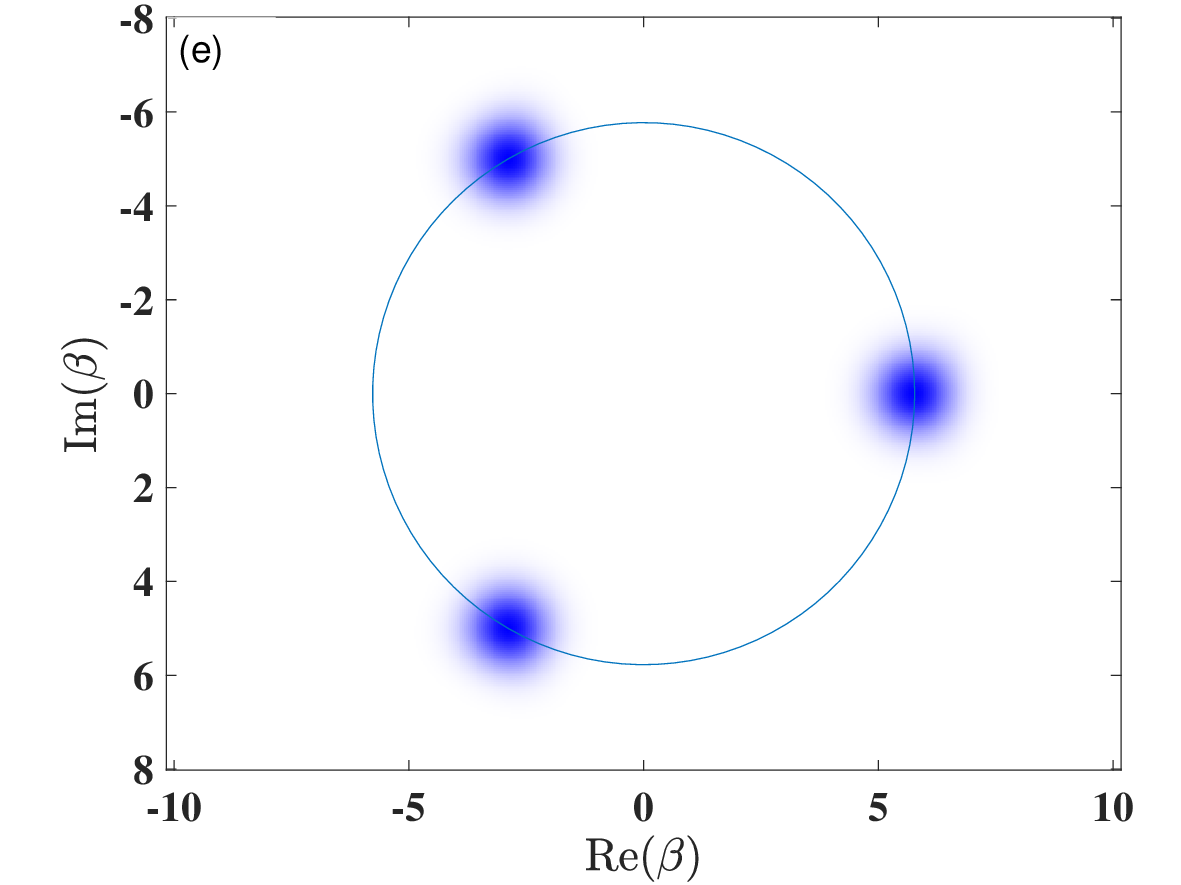}
\includegraphics[width=.3\textwidth]{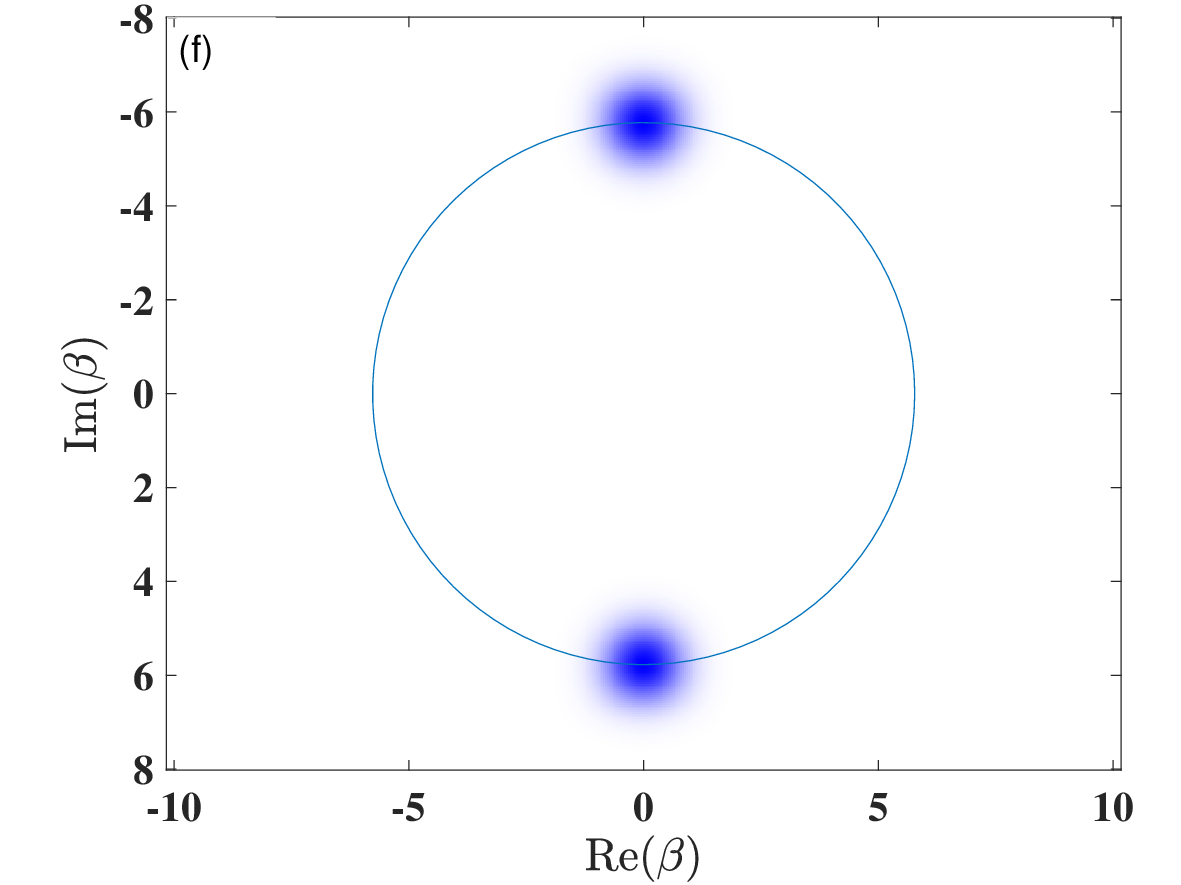}
\caption{The distribution function $|\langle\beta|\alpha\rangle|^M$ with $M=3$ in phase space 
$({\rm Re}(\beta),{\rm Im}(\beta))$ at different times: (a) $Ut=0$, (b) $Ut=\frac{\pi}{6}$, 
(c) $Ut=\frac{2\pi}{6}$, (d) $Ut=\frac{3\pi}{6}$, (e) $Ut=\frac{4\pi}{6}$ and (f) $Ut=\pi$,. 
The characteristic CS parameter is $\alpha=\sqrt{\frac{100}{3}}$.
}\label{fig:circle2}
\end{figure}

\begin{figure}[h]
\centering
\includegraphics[width=0.45\textwidth]
{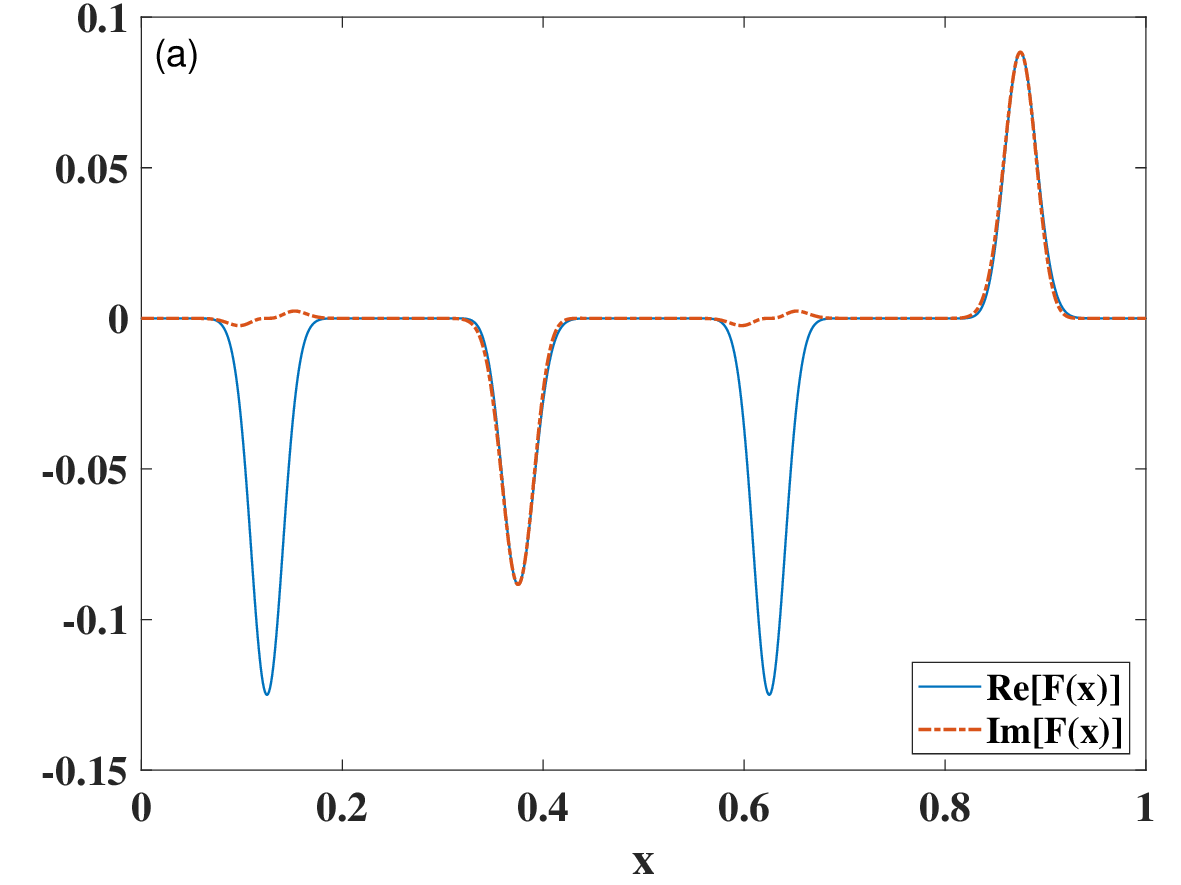}
\includegraphics[width=0.45\textwidth]
{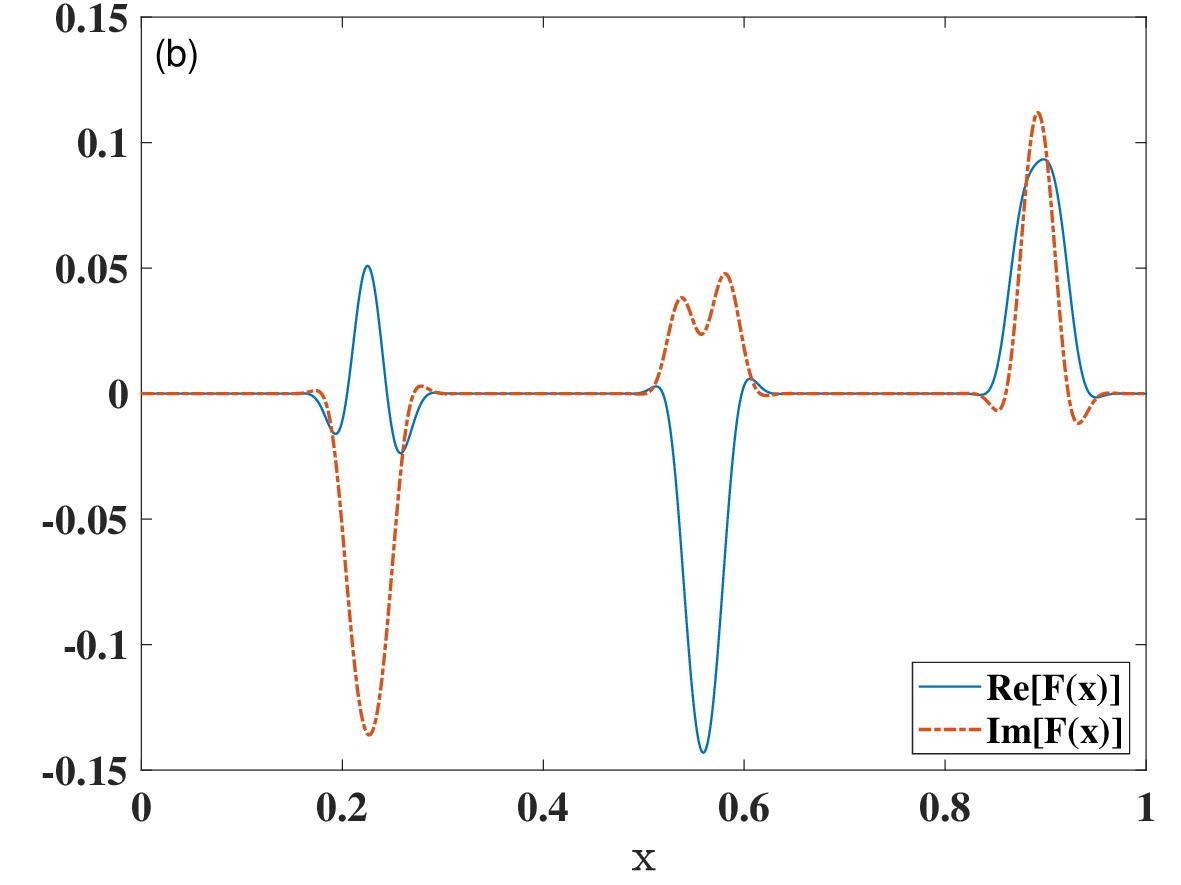}
\caption{The real part (solid blue) and the imaginary part (read dot-dash) of the $F(x,t)$ as the function of $x$ for two fixed time points: (a) $Ut=\frac{\pi}{2}$ and (b) $Ut=2.074$. The system parameters are same with ones in Fig.\ \ref{fig:circle2}.}
\label{fig:Fx}
\end{figure}

In Fig.\ \ref{fig:circle2}, the corresponding distribution function 
$|\langle \beta|{\rm e}^{-{\rm i}\hat{H}t}|\alpha\rangle|^M$ for 
$\alpha=\sqrt{\lambda}=\sqrt{\frac{100}{3}}$ is depicted. This distribution splits 
into several spots which form multi-component cat states at different instances of 
time. To gain some insight, let us focus on the result at the instance $Ut=\pi$, 
displayed in panel (f). The cross-correlation from Eq.\ (\ref{eq:cc}) then reads
\begin{eqnarray}
 c_{\beta\alpha}(\pi/U)&=&
 \exp\left[-\frac{|\alpha|^2+|\beta|^2}{2}\right]\sum_{n=0}^\infty
 \frac{(\alpha\beta^\ast)^n}{n!}{\rm e}^{-{\rm i}\frac{\pi}{2}n(n-1)}
 \nonumber
 \\
 &=&
 \exp\left[-\frac{|\alpha|^2+|\beta|^2}{2}\right]\sum_{n=0}^\infty
 \frac{(\alpha\beta^\ast{\rm e}^{-{\rm i}\frac{\pi}{2}(n-1)})^n}{n!}
 \label{eq:cc}
\end{eqnarray}
and we consider the cases of purely imaginary $\beta_\pm=|\beta|{\rm e}^{\pm{\rm i}\pi/2}$, when the term under the $n$-th power becomes 
$\pm\alpha|\beta|{\rm e}^{-{\rm i}n\pi/2}$. Now using Euler's formula 
${\rm e}^{-{\rm i}n\pi/2}=\cos(n\pi/2)-{\rm i}\sin(n\pi/2)$, we realize that 
the summation over $n$ can be brought into closed form according to
\begin{equation}
\sum_{n=0}^\infty\frac{(\alpha\beta^\ast{\rm e}^{-{\rm i}\frac{\pi}{2}(n-1)})^n}{n!} =\cosh(\alpha|\beta|)\mp{\rm i}\sinh(\alpha|\beta|)
\end{equation}
and for $\alpha=|\beta|\gg 1$, we get
\begin{equation}
    c_{\beta\alpha}(\pi/U)\approx\frac{1}{2}\mp\frac{\rm i}{2}.
\end{equation}
Away from the center of the two spots in panel (f) of Fig. \ref{fig:circle2}, the 
cross-correlation and thus the distribution function will be exponentially suppressed. 
For different special times, as shown in the other panels of Fig.\ \ref{fig:circle2}, 
there may be more of those spots, all of which move 
along the circle with radius $\sqrt{\lambda}$, however, as the quantum 
fluctuation of the Glauber CS with large $\alpha$ is suppressed. Consequently, the 
vanishing of the survival probability can not be attributed to the deviation of 
$G(x,t)$ from the circle, as it was possible before.
\\
Instead, we should examine the integrand on the right hand side of Eq.\ (\ref{eq:fourier})
\begin{eqnarray}
    F(x,t)={\rm e}^{-{\rm i}2\pi xS}G(x,t).
\end{eqnarray} 
In Fig.\ \ref{fig:Fx}, we divide the $F(x,t)$ into real and the imaginary parts, and choose two typical times to observe how they vary with $x$, ranging from $0$ to $1$. In the left figure, the time is fixed at $Ut=\frac{\pi}{2}$ corresponding to Fig.\ \ref{fig:circle2}\ (d) where the four spots manifest as four peaks.
Because three peaks are negative and one is positive, all of them displaying similar heights, the integral of the real part with respect to the $x$ is not zero resulting in a finite survival probability. In the right figure, however, the time is selected to be $Ut=2.074$ identical with the position of the first main spike in Fig.\ \ref{fig:echo2}. In this case, the integral of both real and imaginary components is very small, since the peaks manifest as either positive or negative. The contributions with opposing signs ultimately cancel each other out, giving rise to the decrease of the overall integral. As a result, the survival probability will also go towards zero. In summary, when $\lambda$ is large, the value of the survival probability depends 
on the collective effects of all the spots.

\subsection{General Hamiltonian}

The connection between the auto-correlations of the different types of CS that we have just 
established can be generalized to a generic but number conserving Hamiltonian $\hat{H}$.
A commonly used example would, e.g., be the Bose-Hubbard Hamiltonian, containing the on-site
interaction, we have considered so far as well as an inter-site hopping term, see, e.g., Eq.\ 
(\ref{eq:fb}). To proceed, we recall Eq.\ (\ref{eq:expansion}) and expand the survival 
amplitude of the Glauber CS as
\begin{eqnarray}\label{eq:expansion2}
\langle\vec{\alpha}|{\rm e}^{-{\rm i}\hat{H}t}|\vec{\alpha}\rangle={\rm e}^{-\tilde{N}}\sum_{S'=0}^{\infty}\frac{\tilde{N}^{S'}}{S'!}\langle S',\vec{\xi}|{\rm e}^{-{\rm i}\hat{H}t}|S',\vec{\xi}\rangle.
\end{eqnarray}
The next step is to extract the term $\langle S',\vec{\xi}|{\rm e}^{-{\rm i}\hat{H}t}|S',\vec{\xi}\rangle$ from the sum. An elegant way to perform this task
is to utilize the projection operator onto a specific particle number $S$, as defined in \cite{SDZ11}
\begin{eqnarray}
    \hat{P}=\int^1_0{\rm d}x {\rm e}^{{\rm i}2\pi x(\hat{N}-S)}
\end{eqnarray}
where $\hat{N}=\sum_{i=1}^M a^\dag_i \hat{a}_i$. It is clear that the operator 
$\hat{P}$ plays the role of $\delta_{S',S}$.

Applying the operator $\hat{P}$ to both sides  of Eq.\ (\ref{eq:expansion2}), we get
\begin{eqnarray}
    &\int^1_0{\rm d}x {\rm e}^{-{\rm i}2\pi xS}\langle\vec{\alpha}|{\rm e}^{{\rm i}2\pi x\hat{N}}{\rm e}^{-{\rm i}\hat{H}t}|\vec{\alpha}\rangle\nonumber\\
    &={\rm e}^{-\tilde{N}}\int^1_0{\rm d}x {\rm e}^{-{\rm i}2\pi xS}\sum_{S'=0}^{\infty}\frac{\tilde{N}^{S'}}{S'!}\langle S',\vec{\xi}|{\rm e}^{{\rm i}2\pi x\hat{N}}{\rm e}^{-{\rm i}\hat{H}t}|S',\vec{\xi}\rangle\nonumber\\
    &={\rm e}^{-\tilde{N}}\frac{\tilde{N}^{S}}{S!}\langle S,\vec{\xi}|{\rm e}^{-{\rm i}\hat{H}t}|S,\vec{\xi}\rangle
\end{eqnarray}
which gives rise to the final result
\begin{eqnarray}\label{eq:fourier2}
    \langle S,\vec{\xi}|{\rm e}^{-{\rm i}\hat{H}t}|S,\vec{\xi}\rangle={\rm e}^{\tilde{N}}\frac{S!}{\tilde{N}^{S}}\int^1_0{\rm d}x {\rm e}^{-{\rm i}2\pi xS}\langle\vec{\alpha }{\rm e}^{-{\rm i}2\pi x}|{\rm e}^{-{\rm i}\hat{H}t}|\vec{\alpha}\rangle
\end{eqnarray}
for the auto-correlation in complete analogy to Eq.\ (\ref{eq:fourier}).
By using Stirling's formula and setting $\tilde{N}=S$ as well as 
$\alpha_1=\alpha_2=\cdots=\sqrt{\frac{S}{M}}=\sqrt{\lambda}$, it can be verified that Eq.\ 
(\ref{eq:fourier2}) will reproduce the Eq.\ (\ref{eq:fourier}),
which was derived for the Hamiltonian given in Eq.\ (\ref{hm}). It remains to be seen 
if the general relation can be applied fruitfully in the future.

\section{Conclusions and Outlook}

In the thermodynamic limit, for the interacting boson model,
quantities like two-point correlation functions and the quasi-momentum distribution 
are the same for MMGS and GCS initial conditions.
We have shown here, that for other important quantities, there exist differences 
between those states, even in the thermodynamic limit.
To this end we have studied the quantum mechanical time-evolution of an 
initial wavefunction that was either a MMGS or a GCS and have proven the fact 
that the respective correlations are  related via a Fourier-type relation. 
Pictorially this could be visualized via the overlap of a time evolved Gaussian with a 
set of other Gaussians, whose phases $x$ are distributed along a circle with 
a radius that is defined by the filling factor. All those 
different Gaussians that by themselves are not obeying U(1) symmetry 
correspond to a generalized coherent state, which obeys this symmetry. The Fourier 
coefficient at the value $S$ (corresponding to the fixed particle number) of the 
quantity $G(x,t)$ introduced in the main text is then shown to be the auto-correlation 
function of the generalized coherent state. This function is usually studied in
a typical quench scenario.

Building on the Fourier relation, we could show that there are different physical 
reasons for the correlations to become (almost) zero and thus the free-energy density
to become very large. In the thermodynamic limit, the time evolved state, by moving away 
from the circle, will develop zero overlap at certain instances in time and thus a diverging 
rate function follows, reminiscent of the case of dynamical phase 
transitions \cite{Heyl18}. This connection shall be further explored in the future 
also for the tight-binding model with staggered onsite potential, studied previously
with \cite{RFGS14} and without external flux \cite{FUBSG17}, as well as the full 
Bose-Hubbard model, that has thus far been studied in the context of
quench dynamics using either direct numerical diagonalization \cite{RFGS14}
or matrix product state (DMRG) methods \cite{LaHe19}. Employing generalized 
coherent states, a variational multi-configuration ansatz for the solution of the TDSE, 
along the lines of \cite{pra21}, is then naturally called for.

\section*{Appendix A: Exact analytical solution for the GCS auto-correlation function}

In the following, we explicitly calculate the time evolution of the 
GCS driven by the deep lattice, as defined in Eq.\ (\ref{hm}).
Applying the time-evolution operator, we find
\begin{eqnarray}
 {\rm e}^{-{\rm i}\hat{H}t}|S,\vec{\xi}\rangle\non&={\rm e}^{-{\rm i}\big(\sum_{i=1}^M\hat{h}_i\big)t}\frac{1}{\sqrt{S!}}\Big(\sum_{i=1}^M\xi_i\hat{a}_i^\dag\Big)^S|{\rm vac}\rangle\\
 \non&=\sqrt{S!}\sum_{[n_i]=S}\prod_{i=1}^M\frac{1}{n_i!}{\rm e}^{-{\rm i}\hat{h}_it}(\xi_i\hat{a}_i^\dag)^{n_i}|{\rm vac}\rangle\\
\non&=\sqrt{S!}\sum_{[n_i]=S}\prod_{i=1}^M\frac{1}{n_i!}(\xi_i\hat{a}_i^\dag)^{n_i}{\rm e}^{-{\rm i}\frac{U}{2}(\hat{n}_i+n_i)(\hat{n}_i+n_i-1)t}|{\rm vac}\rangle\\
 \non&=\sqrt{S!}\sum_{[n_i]=S}\prod_{i=1}^M\frac{1}{n_i!}\left[\xi_i{\rm e}^{-{\rm i}\frac{U}{2}(n_i-1)t}\hat{a}_i^\dag\right]^{n_i}|{\rm vac}\rangle\\
 \non&=\sqrt{S!}\sum_{[n_i]=S}{\rm e}^{{\rm i}\frac{U}{2}St}\prod_{i=1}^M\frac{1}{n_i!}\Big(\xi_i{\rm e}^{-{\rm i}\frac{U}{2}n_it}\hat{a}_i^\dag\Big)^{n_i}|{\rm vac}\rangle\\
 &=\sqrt{S!}\sum_{[n_i]=S}\prod_{i=1}^M\frac{1}{n_i!}\Big(\xi_i{\rm e}^{-{\rm i}\frac{U}{2}n_it}\hat{a}_i^\dag\Big)^{n_i}|{\rm vac}\rangle,
\end{eqnarray}
where in the second line we expand the GCS in terms of the Fock states as shown in 
Eq.\ (\ref{eq:GCS_Fock}) and apply the factors ${\rm e}^{-{\rm i}\hat{h}_it}$ 
to the corresponding modes. In the third line we utilize the formula 
\begin{equation}
  \hat{a}_i^\dag f(\hat{n}_i)=f(\hat{n}_i+1)\hat{a}_i^\dag.  
\end{equation} 
The result in the fourth line is given by the fact that
\begin{equation}
    {\rm e}^{-{\rm i}\frac{U}{2}(\hat{n}_i+n_i)(\hat{n}_i+n_i-1)t}|{\rm vac}\rangle={\rm e}^{-{\rm i}\frac{U}{2}n_i(n_i-1)t}|{\rm vac}\rangle,
\end{equation} 
and the global phase factor ${\rm e}^{{\rm i}\frac{U}{2}St}$ in the second last line stems 
from ${\rm e}^{{\rm i}\frac{U}{2}St}=\prod_{i=1}^M {\rm e}^{{\rm i}\frac{U}{2}n_it}$,
which is due to that fact that the sum over all $n_i$ gives the total number of particles, 
see also the text after Eq. (\ref{eq:GCS_Fock}).

The auto-correlation function or Loschmidt amplitude is thus finally given by
\begin{eqnarray}
 \langle S,\vec{\xi}|{\rm e}^{-{\rm i}\hat{H}t}|S,\vec{\xi}\rangle=S!\sum_{[n_i]=S}\prod_{i=1}^M\frac{|\xi_i|^{2n_i}}{n_i!}{\rm e}^{-{\rm i}\frac{U}{2}n_i^2t},
\end{eqnarray}
where the orthogonality of the Fock states was used.

\section*{References}

\bibliographystyle{iopart-num}
\bibliography{ref}

\end{document}